\documentclass[journal]{IEEEtran}   

\usepackage{amsmath}
\usepackage{graphicx,psfrag} 
\usepackage{amsmath,amsthm,amsfonts,amssymb,bm} 
\usepackage[ruled]{algorithm2e}
\usepackage{algpseudocode}
\usepackage{xcolor}
\usepackage{bbm}             
\usepackage{tabularx}        
\usepackage{framed}
\usepackage[ruled]{algorithm2e}
\usepackage[framed,numbered,autolinebreaks,useliterate]{mcode}
\usepackage{cite}
\usepackage[latin1]{inputenc}
\allowdisplaybreaks[3]

\theoremstyle{definition}   
\newcommand{\bs}{\mathbf}
\newtheorem{lemma}{Lemma}
\newtheorem{proposition}{Proposition}
\newtheorem{definition}{Definition}

\newtheorem{assumption}{Assumption}

\newtheorem{remark}{Remark}

\begin{document}

\title{
Replica Analysis for Generalized Linear Regression with IID Row Prior
}

\author{
Qiuyun Zou, and Hongwen Yang

\thanks{
Q. Zou and H. Yang are with Beijing University of Posts and Telecommunications,
Beijing 100876, China (email: qiuyun.zou@bupt.edu.cn; yanghong@bupt.edu.cn).
}
}
\maketitle

\begin{abstract}
Different from a typical independent identically distributed (IID) element assumption, this paper studies the estimation of IID row random matrix for the generalized linear model constructed by a linear mixing space and a row-wise mapping channel. The objective inference problem arises in many engineering fields, such as wireless communications, compressed sensing, and phase retrieval. We apply the replica method from statistical mechanics to analyze the exact minimum mean square error (MMSE) under the Bayes-optimal setting, in which the explicit replica symmetric solution of the exact MMSE estimator is obtained. Meanwhile, the input-output mutual information relation between the objective model and the equivalent single-vector system is established. To estimate the signal, we also propose a computationally efficient message passing based algorithm on expectation propagation (EP) perspective and analyze its dynamics. We verify that the asymptotic MSE of proposed algorithm predicted by its state evolution (SE) matches perfectly the exact MMSE estimator predicted by the replica method. That indicates, the optimal MSE error can be attained by the proposed algorithm if it has a unique fixed point.
\end{abstract}


\section{Introduction}
We consider high-dimensional signal recovery of $\bs{X}$ from a matrix observation $\bs{Y}$ for the generalized linear regression:
\begin{align}
\mathcal{P}(\bs{Y}|\bs{Z})=\prod_{\ell=1}^L \mathcal{P}(\bs{y}_{\ell}|\bs{z}_{\ell}), \ \text{s.t.}, \ \bs{Z}=\bs{HX},
\label{Equ:objective}
\end{align}
where $\bs{H}\in \mathbb{C}^{L\times N}$ is known as measurement matrix, $\bs{X}\in \mathbb{C}^{N\times M}$ is the signal of interest whose each row entry is assumed to be drawn from a random vector $\textsf{X}$, and the observation $\bs{Y}$ is from $\bs{Z}=\bs{HX}$ through a row-wise mapping channel $\mathcal{P}(\bs{Y}|\bs{Z})=\prod_{\ell=1}^L \mathcal{P}(\bs{y}_{\ell}|\bs{z}_{\ell})$ with $\bs{y}_{\ell}\in \mathbb{C}^{M}$ and $\bs{z}_{\ell}\in \mathbb{C}^M$ being the $\ell$-th row of $\bs{Y}$ and $\bs{Z}$, respectively. This inference problem arises in many engineering domains, such as wireless communications \cite{liu2018massive,liu2019generalized,zou2020message}, phase retrieval \cite{wang2020decentralized,schniter2014compressive,ma2021spectral}, signal detection \cite{zou2020low}, etc.  In \cite{liu2018massive,liu2019generalized}, two approximate message passing (AMP) \cite{donoho2009message} based methods were developed for the joint channel and user activity estimation considering massive access scenario, in which each row of $\bs{X}$ was assumed to be drawn from Bernoulli-Gaussian distribution (identity covariance matrix Gaussian component). A novel modeling for such massive access scenario was proposed in \cite{zou2020message} by separating the activity factor from the compound channel, and a hybrid algorithm combining loopy belief propagation  (LBP) and AMP was proposed based on the new modeling which can achieve the same performance as \cite{liu2018massive}  but with lower computational cost. In \cite{wang2020decentralized}, the authors proposed a decentralized expectation consistent (EC) framework to address phase retrieval problem. In fact, all of them are special cases of the objective inference problem by specifying priors and transition distributions.

The main methods used in this paper refer to message passing and replica method.  Message passing algorithms have attracted a lot of research attention since the Donoho's AMP \cite{donoho2009message}. The AMP algorithm was originally proposed to provide sparse solution for the least absolute shrinkage and selection operator (LASSO) problem. The dynamics of AMP was mathematically analyzed in \cite{bayati2011dynamics} in the large system limit, where the mean square error (MSE) performance of AMP can be fully tracked by a scalar iteration termed state evolution (SE). Following AMP, \cite{rangan2011generalized} extended AMP to the generalized linear model and proposed generalized AMP (GAMP) which allows arbitrary element-wise mapping channel. Another algorithm related to AMP refers to expectation propagation (EP) \cite{minka2001family} which is very close to EC \cite{opper2005expectation} (single loop version), orthogonal AMP (OAMP) \cite{ma2017orthogonal}, and vector AMP (VAMP) \cite{rangan2019vector}. They were proposed independently, but it is easy to verify that they share the same algorithm after singular value decomposition (SVD). Recently, some works \cite{zou2018concise,meng2015concise} show that AMP has a strong correlation to EP, in which the EP is applied to provide a concise derivation of AMP. The analysis tool applied in this paper is the replica method from statistical physics \cite{mezard1987spin}. The replica method is known as a heuristic method, but the results using this method are widely verified to be exact \cite{kabashima2016phase,rangan2012asymptotic,barbier2019adaptive,reeves2019replica,schulke2016statistical, vehkapera2016analysis,bereyhi2019statistical,zou2021multi}.
In \cite{barbier2019adaptive}, a rigorous method called adaptive interpolation method was applied to prove the replica formula. The recent work \cite{reeves2019replica} also proved that the replica symmetric solution is correct in the case of independent identically distributed (IID) Gaussian matrices.  In \cite{tanaka2002statistical,guo2005randomly}, the replica method was applied to analyze the dynamics of code-division multiple-access (CDMA) multiuser detector, where the replica symmetric solution is the same as the SE of AMP \cite{donoho2010message}. For some basic knowledge of replica method and its applications, see \cite{zdeborova2016statistical}.

The objective inference problem is an extension of the generalized linear model \cite{rangan2011generalized} by considering the correlations in both the signal of interest and transition mapping channel. The main focus here, besides the proposed algorithm with detailed derivation using EP, is the asymptotic analysis for the exact minimum mean square error (MMSE) estimator under the Bayes-optimal setting which was not studied in \cite{rangan2011generalized}. The replica analysis presented here can cover the case in \cite{rangan2011generalized} by specifying IID element priors and element-wise transition distributions. The contributions of this paper are summarized as follows:
\begin{itemize}
\item Different from previous IID element assumption, we here consider IID row prior and row-wise mapping channel. We give the replica analysis for the objective generalized linear regression inference problem, which provides explicit formulas for the exact MMSE estimator in the Bayes-optimal setting and large system limit. Meanwhile, the input-output mutual information relation between the objective model and the equivalent single-vector system is established.
\item We propose a computational efficient message passing based algorithm for estimating $\bs{X}$ for the objective inference problem. We present the detailed derivation of proposed algorithm on EP perspective, which unifies the complex-valued and real-valued regions.
\item In addition, by performing the large system analysis, we verify that the asymptotic MSE of proposed algorithm predicted by its SE matches  perfectly the achievable MMSE predicted by replica method, which is well-known to be Bayes-optimal but computational disastrous.
\end{itemize}

The remainder of this work is organized as follows. Section II states the objective problem. Section III introduces the main results of this paper. Section IV gives the replica analysis to analyze the achievable MMSE in Bayes-optimal setting and large system limit.  Section V presents the proposed algorithm and analyzes its asymptotic MSE performance. Finally, Section VI gives several numerical simulations to validate the accuracy of our analytical results.

\textbf{Notations:} We use $\bs{x}$ and $\bs{X}$ to denote column vector and matrix. $(\cdot)^{\text{T}}$ refers to transpose operation while $(\cdot)^{\dag}$ refers to Hermitian transpose operation. $\text{Tr}(\bs{X})$ denotes the trace of $\bs{X}$. $\det(\bs{X})$ denotes the determinant of $\bs{X}$.
$\Re(\bs{x})$ denotes the real part of $\bs{x}$ while $\Im(\bs{x})$ denotes the image part of $\bs{x}$.
$\otimes $ denotes Kronecker product. $\bs{x}^{\times i}$ denotes $i$ Kronecker product of $\bs{x}$, such as $\bs{x}^{\times 2}=\bs{x}\otimes \bs{x}$. $I(\cdot)$ is mutual information. $\mathbb{E}_{\bs{x}}\{\cdot\}$ denotes the expectation over $\mathcal{P}(\bs{x})$. We use $\textsf{MMSE}_{\bs{X}}$ and $\textsf{MSE}_{\bs{X}}$ to denote MMSE matrix and its MSE for the exact MMSE estimator, and
$\textsf{MMSE}(\bs{X},t)$ and $\textsf{MSE}(\bs{X},t)$  to denote asymptotic MMSE matrix and its MSE of the proposed algorithm at $t$-iteration, respectively.
$\mathcal{N}_c(\bs{x}|\bs{a},\bs{A})$ is complex-valued vector Gaussian
with argument $\bs{x}$, $\bs{a}$ mean and $\bs{A}$ covariance matrix $\mathcal{N}_c(\bs{x}|\bs{a}, \bs{A})=\det(\pi\bs{A})^{-1}\exp \left[-(\bs{x}-\bs{a})^{\dag}\bs{A}^{-1}(\bs{x}-\bs{a})\right].$ $\mathcal{N}_c(\bs{a},\bs{A})$ denotes a complex-valued vector Gaussian with $\bs{a}$ mean and $\bs{A}$ covariance matrix.

\begin{figure}[!t]
\centering
\includegraphics[width=0.45\textwidth]{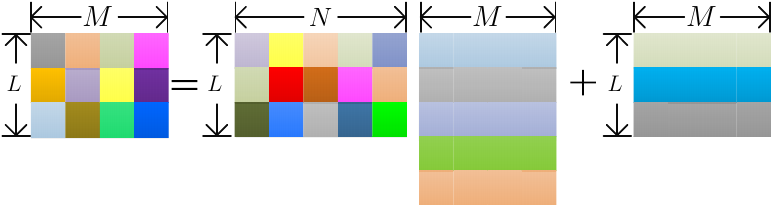}
\caption{A simple example $\bs{Y}=\bs{HX}+\bs{W}$ with dimensions $(L,N,M)=(3, 5, 4)$ of objective model (\ref{Equ:objective}), where each row of $\bs{X}$ follows $\mathcal{P}_{\textsf{X}}$ and each row of $\bs{W}$ follows correlated Gaussian $\mathcal{N}_c(\bs{0},\bs{\Sigma}_w)$. But, be aware, our focus is the large system limit.
}
\label{fig:AWGN}
\end{figure}

\section{Problem Formalization}
\subsection{System Model}
Consider the signal recovery of $\bs{X}\in \mathbb{C}^{N\times M}$ from $\bs{Y}\in \mathbb{C}^{L\times M}$ with the form:
\begin{align}
\mathcal{P}(\bs{Y}|\bs{Z})=\prod_{\ell=1}^L\mathcal{P}(\bs{y}_{\ell}|\bs{z}_{\ell}) \  \ \text{s.t.},\ \bs{Z}=\bs{HX},
\label{Equ:system}
\end{align}
where the observation $\bs{Y}$ is from $\bs{X}$ passing a linear mixing space $\bs{Z}=\bs{HX}$ and a row-wise mapping channel $\mathcal{P}(\bs{Y}|\bs{Z})=\prod_{\ell=1}^L\mathcal{P}(\bs{y}_{\ell}|\bs{z}_{\ell})$ with $\bs{y}_{\ell}\in \mathbb{C}^M$ and $\bs{z}_{\ell}\in \mathbb{C}^M$ being the $\ell$-th row of $\bs{Y}$ and $\bs{Z}$, respectively. In doing so, it is assumed that each row $\bs{x}_n\in \mathbb{C}^M$ of $\bs{X}$ follows IID random vector (RV) $\textsf{X}$ drawn from the known probability distribution $\mathcal{P}_{\textsf{X}}$, i.e.,
\begin{align}
\mathcal{P}(\bs{X})=\prod_{n=1}^N \mathcal{P}_{\textsf{X}}(\bs{x}_n).
\end{align}
As a simple example, Fig.\ref{fig:AWGN} shows the framework of the objective system under correlated Gaussian transition. Also, we assume that $\textsf{X}$ is a RV with zero mean and $\bs{\Xi}_x$ covariance matrix. $\bs{H}\in \mathbb{C}^{L\times N}$ is the measurement matrix which is known beforehand and each element of $\bs{H}$ is drawn from a random variable with zero mean and $1/L$ variance. Throughout, we focus on the \textit{Bayes-optimal setting}, where both prior and likelihood function are precisely known beforehand. Also, we consider the \textit{large system limit}, in which the dimensions of system tend to infinite $L,N\uparrow \infty$ but the ratio $\alpha=\frac{L}{N}$ is bounded and fixed.

\subsection{MMSE Estimator}
In Bayes-optimal setting, the MMSE estimator \cite[Chapter 11]{kay1993fundamentals}, also known as posterior mean estimator (PME), is optimal in MSE sense. Then, we treat the objective inference problem into Bayesian inference framework. The MMSE estimator of $\bs{X}$ is given by
\begin{align}
\forall n: \ \hat{\bs{x}}_n=\mathbb{E}\left\{\bs{x}_n|\bs{Y},\bs{H}\right\},
\label{Equ:Marginal}
\end{align}
where the distribution is taken over the marginal posterior $\mathcal{P}(\bs{x}_n|\bs{Y},\bs{H})$ which is obtained by marginalizing $\mathcal{P}(\bs{X}|\bs{Y},\bs{H})$. By Bayes rules, the posterior can be represented as
\begin{align}
\mathcal{P}(\bs{X}|\bs{Y},\bs{H})
&=\frac{\mathcal{P}(\bs{Y}|\bs{H},\bs{X})\mathcal{P}(\bs{X})}{\mathcal{P}(\bs{Y}|\bs{H})}.
\end{align}

The MMSE matrix corresponding to MMSE estimator $\hat{\bs{X}}=\{\hat{\bs{x}}_n,\forall n\}$ is represented as
\begin{align}
\!\!\!\!\!\!\textsf{MMSE}_{\bs{X}}=\mathbb{E}\left\{(\bs{X}-\mathbb{E}\left\{\bs{X}|\bs{Y},\bs{H}\right\})(\bs{X}-\mathbb{E}\left\{\bs{X}|\bs{Y},\bs{H}\right\})^{\dag}\right\}
\end{align}
The averaged MSE associated with MMSE matrix is defined by
\begin{align}
\textsf{MSE}_{\bs{X}}
&=\frac{1}{NM}\text{Tr}(\textsf{MMSE}_{\bs{X}}).
\end{align}
Notice that, there exists $(N-1)M$ integrals to obtain $\mathcal{P}(\bs{x}_n|\bs{Y},\bs{H})$ in (\ref{Equ:Marginal}) from $\mathcal{P}(\bs{X}|\bs{Y},\bs{H})$. The computational cost of this marginalization is extremely prohibitive except the simplest case such as both prior and transition distribution being Gaussian. To avoid this computational disaster, it is inevitable to make some approximations  to  MMSE estimator. The first is how to analyze the achievable MSE performance of the exact MMSE estimator. The second is how to design a practical algorithm whose asymptotic MSE performance achieves the Bayes-optimal MSE performance. For the first issue, we apply the replica method from statistical physics to analyze the exact MMSE estimator, which provides a benchmark for designing a new algorithm. For the second issue, we develop a low-complexity message passing based algorithm on expectation propagation perspective, whose computational cost is $\mathcal{O}(N^2+M^2N)$ and it can attain Baeys-optimal MSE performance if its fixed point is unique.

\section{Main Results}
\begin{figure}[!t]
\centering
\includegraphics[width=0.3\textwidth]{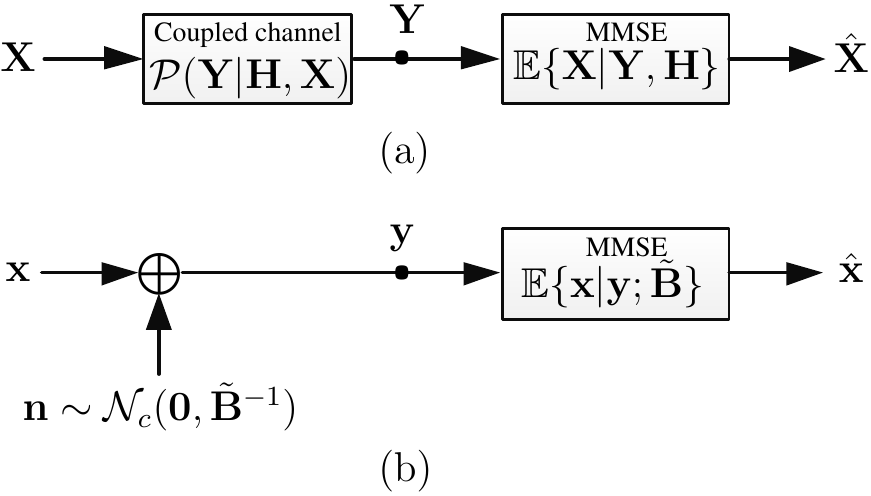}
\caption{(a) The multiple-input multiple-output system and MMSE estimator; (b) the equivalent single-vector Gaussian channel and MMSE estimator, where the parameter $\tilde{\bs{B}}$ implicitly involves the mean and variance of $\bs{H}$.
}
\label{fig:SISO}
\end{figure}

This section gives the main results of this paper considering complex-valued generalized linear model with IID row prior and row-wise mapping channel. All of those results are for \textit{large system limits}. That is, the dimensions of system tend to infinite $L,N\uparrow \infty$ but the ratio $\alpha=\frac{L}{N}$ is bounded and fixed. The detailed procedures are respectively given in Section \ref{Sec:RM} and Section \ref{Sec:MP}.

\subsection{Equivalent Single-Vector Channel}
To derive main results, we first introduce the following equivalent single-vector channel
\begin{align}
\bs{y}=\bs{x}+\bs{n},
\label{Equ:vecchannel}
\end{align}
where $\bs{x}\in \mathbb{C}^{M}$ is distributed as $\mathcal{P}_{\textsf{X}}$ with zero mean and $\bs{\Xi}_x$ covariance matrix, and $\bs{n}$ is complex additive white Gaussian noise with $\mathcal{N}_c(\bs{0},\tilde{\bs{B}}^{-1})$, where $\tilde{\bs{B}}$ is from the fixed point equations of the exact MMSE estimator and will be introduced in the rest of this section. Accordingly, the likelihood function is given by
\begin{align}
\!\!\!\!\mathcal{P}(\bs{y}|\bs{x};\tilde{\bs{B}})=\det(\pi \tilde{\bs{B}}^{-1})^{-1}\exp \left[-(\bs{y}-\bs{x})^{\dag}\tilde{\bs{B}}(\bs{y}-\bs{x})\right].
\end{align}

The MMSE estimator of $\bs{x}$ denotes
\begin{align}
\hat{\bs{x}}=\mathbb{E}_{\bs{x}|\bs{y}}\{\bs{x}\}=\int \bs{x}\mathcal{P}(\bs{x}|\bs{y})\text{d}\bs{x},
\end{align}
where
$
\mathcal{P}(\bs{x}|\bs{y})=\frac{\mathcal{P}_{\textsf{X}}(\bs{x})\mathcal{P}(\bs{y}|\bs{x};\tilde{\bs{B}})}{\mathcal{P}(\bs{y})}
$ according to the Bayes rules.
The MMSE matrix of the MMSE estimator $\hat{\bs{x}}$ refers to
\begin{align}
\nonumber
\textsf{MMSE}_{\bs{x}}
&=\mathbb{E}_{\bs{x},\bs{y}}\left\{(\bs{x}-\mathbb{E}\{\bs{x}|\bs{y}\})(\bs{x}-\mathbb{E}\{\bs{x}|\bs{y}\})^{\dag}\right\}\\
&=\bs{A}-\bs{B},
\label{Equ:MSE}
\end{align}
where $\bs{A}=\mathbb{E}_{\bs{x}}\{\bs{x}\bs{x}^{\dag}\}$ and $\bs{B}=\mathbb{E}_{\bs{y}}\{\mathbb{E}_{\bs{x}|\bs{y}}\{\bs{x}\}\mathbb{E}_{\bs{x}|\bs{y}}\{\bs{x}\}^{\dag}\}$. The MSE associated with MMSE matrix is given by
\begin{align}
\textsf{MSE}_{\bs{x}}=\frac{1}{M}\text{Tr}(\textsf{MMSE}_{\bs{x}}).
\end{align}

In addition, the input-output mutual information $I(\bs{x};\bs{y})$ of (\ref{Equ:vecchannel}) denotes
\begin{align}
\nonumber
I(\bs{x};\bs{y})
&=\int_{\bs{y}}\int_{\bs{x}}\mathcal{P}(\bs{x},\bs{y})\log \frac{\mathcal{P}(\bs{y}|\bs{x};\tilde{\bs{B}})}{\mathcal{P}(\bs{y})}\text{d}\bs{x}\text{d}\bs{y}\\
\nonumber
&=-\mathbb{E}_{\bs{y}}\left\{\log \mathcal{P}(\bs{y})\right\}+\mathbb{E}_{\bs{y}, \bs{x}}\left\{\log \mathcal{P}(\bs{y}|\bs{x};\tilde{\bs{B}})\right\}\\
&=-\mathbb{E}_{\bs{y}}\left\{\log \mathcal{P}(\bs{y})\right\}-M\log \pi+\log \det(\tilde{\bs{B}})-M.
\end{align}

\subsection{Analytical Results}
\begin{proposition}[Decoupling principle]
\label{Pro:Decoupling}
In large system limit, the joint distribution of $(\bs{x}_n, \hat{\bs{x}}_n)$ of a  coupled multiple-input and multiple-output system (\ref{Equ:system}) as depicted in Fig.~\ref{fig:SISO}(a) converges to the joint distribution of $(\bs{x}, \hat{\bs{x}})$ of the equivalent single-vector channel in (\ref{Equ:vecchannel}) as depicted in Fig.~\ref{fig:SISO}(b), i.e.,
\begin{align}
\lim_{N\uparrow \infty} \mathcal{P}(\bs{x}_n, \hat{\bs{x}}_n)=\mathcal{P}(\bs{x},\hat{\bs{x}}).
\end{align}
The parameter $\tilde{\bs{B}}$ in (\ref{Equ:vecchannel}) satisfies the coupled equations
\begin{subequations}
\begin{align}
\bs{B}_z&=\int \frac{\bs{U}_{\bs{y}|\bs{z}}\left(\frac{1}{\sqrt{\alpha}}\bs{B}^{\frac{1}{2}}\boldsymbol{\zeta}, \frac{1}{\alpha}(\bs{A}-\bs{B})\right)}
{v_{\bs{y}|\bs{z}}\left(\frac{1}{\sqrt{\alpha}}\bs{B}^{\frac{1}{2}}\boldsymbol{\zeta}, \frac{1}{\alpha}(\bs{A}-\bs{B})\right)}\text{\rm{D}}\boldsymbol{\zeta}{\rm{d}}\bs{y},\\
\tilde{\bs{B}}&=(\bs{A}-\bs{B})^{-1}(\alpha^2\bs{B}_z-\alpha\bs{B})(\bs{A}-\bs{B})^{-1},\\
\bs{A}&=\bs{\Xi}_x,\\
\bs{B}&=\int \frac{\bs{U}_{\bs{x}}(\boldsymbol{\zeta}, \tilde{\bs{B}}^{-1})}{v_{\bs{x}}(\boldsymbol{\zeta}, \tilde{\bs{B}}^{-1})}{\rm{d}}\boldsymbol{\zeta},
\label{SD:def4}
\end{align}
\label{SD}
\end{subequations}
where $\bs{A}$, $\bs{B}$, and $\tilde{\bs{B}}$ are $M\times M$ real-valued positive definite matrices, $\boldsymbol{\zeta}\in \mathbb{C}^{M}$, and $\text{D}\boldsymbol{\zeta}=\mathcal{N}_c(\boldsymbol{\zeta}|\bs{0},\bs{I})\text{d}\boldsymbol{\zeta}$. In addition, the following definitions are used
\begin{subequations}
\begin{align}
v_{\bs{y}|\bs{z}}(\bs{a},\bs{A})&=\int \mathcal{P}(\bs{y}|\bs{z})\mathcal{N}_c(\bs{z}|\bs{a},\bs{A}){\rm{d}}\bs{z},\\
\bs{u}_{\bs{y}|\bs{z}}(\bs{a}, \bs{A})&=\int \bs{z}\mathcal{P}(\bs{y}|\bs{z})\mathcal{N}_c(\bs{z}|\bs{a},\bs{A}){\rm{d}}\bs{z},\\
\bs{U}_{\bs{y}|\bs{z}}(\bs{a}, \bs{A})&=\bs{u}_{\bs{y}|\bs{z}}(\bs{a}, \bs{A})\bs{u}_{\bs{y}|\bs{z}}(\bs{a}, \bs{A})^{\dag},\\
v_{\bs{x}}(\bs{a},\bs{A})&=\int \mathcal{P}_{\textsf{X}}(\bs{x})\mathcal{N}_c(\bs{x}|\bs{a}, \bs{A}){\rm{d}}\bs{x},\\
\bs{u}_{\bs{x}}(\bs{a},\bs{A})&=\int \bs{x}\mathcal{P}_{\textsf{X}}(\bs{x})\mathcal{N}_c(\bs{x}|\bs{a},\bs{A}){\rm{d}}\bs{x},\\
\bs{U}_{\bs{x}}(\bs{a},\bs{A})&=\bs{u}_{\bs{x}}(\bs{a},\bs{A})\bs{u}_{\bs{x}}(\bs{a}, \bs{A})^{\dag}.
\end{align}
\label{Equ:def}
\end{subequations}
If the solution of (\ref{SD}) is not unique, then the parameter $\tilde{\bs{B}}$ is chosen by minimizing the free energy
\begin{align}
\nonumber
\mathcal{F}
&=-\frac{1}{N}\mathbb{E}_{\bs{Y},\bs{H}}\left\{\log \mathcal{P}(\bs{Y}|\bs{H})\right\}\\
\nonumber
&= -\left(\mathbb{E}_{\bs{y}}\left\{\log \mathcal{P}(\bs{y})\right\}+M\log \pi-\log \det(\tilde{\bs{B}})+M\right)\\
\nonumber
&\quad -\text{Tr}((\bs{A}-\bs{B})\tilde{\bs{B}})
-\alpha \int v_{\bs{y}|\bs{z}}\left(\frac{1}{\sqrt{\alpha}}\bs{B}^{\frac{1}{2}}\boldsymbol{\zeta}, \frac{1}{\alpha}(\bs{A}-\bs{B})\right)\\
&\qquad \times \log v_{\bs{y}|\bs{z}}\left(\frac{1}{\sqrt{\alpha}}\bs{B}^{\frac{1}{2}}\boldsymbol{\zeta}, \frac{1}{\alpha}(\bs{A}-\bs{B})\right)\text{D}\boldsymbol{\zeta}\text{d}\bs{y}.
\label{FreeEnergy}
\end{align}
\proof See Section \ref{Sec:RM}.
\end{proposition}

\begin{remark}
Proposition \ref{Pro:Decoupling} shows that, in high-dimensional system, the input-output of each row $\bs{x}_n$ of the system (\ref{Equ:system}) using the MMSE estimator is precisely decoupled into a single-vector channel (\ref{Equ:vecchannel}). This property is called decoupling principle. The equivalence is that the joint distribution $\mathcal{P}(\bs{x}_n, \hat{\bs{x}}_n)$ of high-dimensional system (\ref{Equ:system}) with arbitrary IID row input and row-wise transitions converges to the joint distribution $\mathcal{P}(\bs{x}, \hat{\bs{x}})$ of the equivalent single-vector channel. The decoupling property can be proved by the equivalent joint moments, i.e., $\mathbb{E}\left\{\Re(\bs{x}_n)^{\times i_{\text{R}}}\otimes \Im(\bs{x}_n)^{\times i_{\text{I}}}\otimes \Re(\hat{\bs{x}}_n)^{\times j_{\text{R}}}\otimes \Im(\hat{\bs{x}}_n)^{\times j_{\text{I}}}
\right\}
=\mathbb{E}\left\{\Re(\bs{x})^{\times i_{\text{R}}}\otimes \Im(\bs{x})^{\times i_{\text{I}}}\otimes \Re(\hat{\bs{x}})^{\times j_{\text{R}}}\otimes \Im(\hat{\bs{x}})^{\times j_{\text{I}}} \right\}$, where $i_{\text{R}}$, $i_{\text{I}}$, $j_{\text{R}}$, and $j_{\text{I}}$ are non-negative integers. The detailed proof of the equivalent joint moments using replica method is presented in Section \ref{Sec:RM}. Specially, it is naturally obtaining the equivalent MMSE matrix, i.e., $\lim_{N\rightarrow \infty}\mathbb{E}\{(\bs{x}_n-\hat{\bs{x}}_n)(\bs{x}-\hat{\bs{x}}_n)^{\dag}\}=\mathbb{E}\{(\bs{x}-\hat{\bs{x}})(\bs{x}-\hat{\bs{x}})^{\dag}\}$. It further implies the equivalent MSE. In addition, the MSE of the equivalent single-vector channel can be determined once $\tilde{\bs{B}}$ is given which is from the fixed point equations (\ref{SD}).  In other words, the fixed point equations (\ref{SD}) predicted by the replica method provide a benchmark that measures whether a practical algorithm achieve Bayes-optimal MSE performance.
\end{remark}

Apart from the replica analysis for the exact MMSE estimator, we develop a message passing based algorithm using EP. Similar to other message passing-like algorithms, we give the large system limit (LST) analysis to analyze the asymptotic MSE performance of proposed algorithm. From those procedures, we have the following results.

\begin{proposition}
In large system limit, by performing the SE analysis, the asymptotic MSE of proposed algorithm can be fully tracked by a set of iterative equations termed as SE. Compared to the fixed point equations (\ref{SD}) of the exact MMSE estimator derived from replica method, SE of proposed algorithm matches perfectly the fixed point (\ref{SD}) of the exact MMSE estimator as predicted by replica method. This consistency indicates that the proposed algorithm could achieve the Bayes-optimal MSE if it has a unique fixed point.
\proof See Section \ref{Sec:MP}.
\end{proposition}

\subsection{Discussion}
The input-output mutual information\footnote{
Throughout, the mutual information of system (\ref{Equ:system}) refers to the averaged mutual information over $\bs{H}$. However, in large system limit, due to self-averaging property \cite{guo2005randomly}, the mutual information for arbitrary realization $\bs{H}$ converges to its expectation over $\bs{H}$.
} of system (\ref{Equ:system}) is given by
\begin{align}
\nonumber
\!\!\!\!\!I(\bs{X};\bs{Y}|\bs{H})
&=\int \mathcal{P}(\bs{X},\bs{Y},\bs{H})\log \frac{\mathcal{P}(\bs{Y}|\bs{H},\bs{X})}{\mathcal{P}(\bs{Y}|\bs{H})}\text{d}\bs{X}\text{d}\bs{Y}\text{d}\bs{H}\\
&=N\mathcal{F}-\mathbb{E}\left\{\mathcal{H}(\bs{Y}|\bs{H},\bs{X})\right\},
\label{Equ:mutual}
\end{align}
where the first term is related to free energy defined in (\ref{FreeEnergy}) and the second term is the expectation of conditional entropy $\mathcal{H}(\bs{Y}|\bs{H},\bs{X})=-\mathbb{E}_{\bs{Y}|\bs{H},\bs{X}}\left\{\log \mathcal{P}(\bs{Y}|\bs{H},\bs{X})\right\}$, which is determined once the transition distribution $\mathcal{P}(\bs{Y}|\bs{Z})$ is specified.

Substituting (\ref{FreeEnergy}) into (\ref{Equ:mutual}), the input-output mutual information relation between the high-dimensional system and  the equivalent single-vector channel can be established as
\begin{align}
\nonumber
&I(\bs{X};\bs{Y}|\bs{H})=NI(\bs{x};\bs{y})-N\text{Tr}((\bs{A}-\bs{B})\tilde{\bs{B}})\\
\nonumber
&\quad -\mathbb{E}\left\{\mathcal{H}(\bs{Y}|\bs{H},\bs{X})\right\}-L\int v_{\bs{y}|\bs{z}}\left(\frac{1}{\sqrt{\alpha}}\bs{B}^{\frac{1}{2}}\boldsymbol{\zeta}, \frac{1}{\alpha}(\bs{A}-\bs{B})\right)\\
&\quad \times \log v_{\bs{y}|\bs{z}}\left(\frac{1}{\sqrt{\alpha}}\bs{B}^{\frac{1}{2}}\boldsymbol{\zeta}, \frac{1}{\alpha}(\bs{A}-\bs{B})\right)\text{D}\boldsymbol{\zeta}\text{d}\bs{y}.
\label{Mutual}
\end{align}
Obviously, once the transition distribution $\mathcal{P}(\bs{Y}|\bs{Z})$ is specified, the mutual information $I(\bs{X};\bs{Y}|\bs{H})$ can be constructed by $I(\bs{x};\bs{y})$ and loss terms, where the parameters $\bs{B}$ and $\tilde{\bs{B}}$ are from the fixed point equations in (\ref{SD}).

\begin{remark}
Actually, the Proposition \ref{Pro:Decoupling} and (\ref{Mutual}) are general, since they allow arbitrary IID row prior and row-wise mapping channel. We now show the degenerated results under the row-wise Gaussian transition distribution $\mathcal{P}(\bs{y}_{\ell}|\bs{z}_{\ell})=\mathcal{N}_c(\bs{y}_{\ell}|\bs{z}_{\ell}, \bs{\Sigma}_w)$. In this case, by Gaussian reproduction property\footnote{
$\mathcal{N}_c(\bs{x}|\bs{a},\bs{A})\mathcal{N}_c(\bs{x}|\bs{b},\bs{B})=\mathcal{N}_c(\bs{0}|\bs{a}-\bs{b}, \bs{A}+\bs{B})\mathcal{N}(\bs{x}|\bs{c}, \bs{C})$ with $\bs{C}=(\bs{A}^{-1}+\bs{B}^{-1})^{-1}$ and $\bs{c}=\bs{C}(\bs{A}^{-1}\bs{a}+\bs{B}^{-1}\bs{b})$.
}, the fixed point equations in (\ref{SD}) reduces as
\begin{align}
\tilde{\bs{B}}^{-1}&=\bs{\Sigma}_w+\frac{1}{\alpha}\textsf{MMSE}_{\bs{x}},
\label{FD:vector}
\end{align}
where $\textsf{MMSE}_{\bs{x}}=\bs{A}-\bs{B}$ defined in (\ref{Equ:MSE}) is the MMSE matrix of the equivalent single-vector channel (\ref{Equ:vecchannel}). In addition, substituting $\mathcal{P}(\bs{y}_{\ell}|\bs{z}_{\ell})=\mathcal{N}_c(\bs{y}_{\ell}|\bs{z}_{\ell}, \bs{\Sigma}_w)$ and (\ref{FD:vector}) into (\ref{Mutual}), we obtain
\begin{align}
\nonumber
I(\bs{X};\bs{Y}|\bs{H})
&=NI(\bs{x};\bs{y})+L\text{Tr}(\tilde{\bs{B}}\bs{\Sigma}_w)\\
&\quad -L\log \det(\bs{\tilde{\bs{B}}}\bs{\Sigma}_w)-LM,
\end{align}
 Considering $M=1$ and $\mathcal{N}_c(y_{\ell}|z_{\ell}, \Sigma_w)$, it is not difficult to observe that the fixed point of exact MMSE estimator as predicted by replica method reduces to $\tilde{B}^{-1}=\Sigma_w+\frac{1}{\alpha}\textsf{MSE}_x$, which coincides the result of \cite{guo2005randomly}. On the other hand, that is also independently found as the SE of the celebrated AMP algorithm \cite{bayati2011dynamics}. Meanwhile, we also get $I(\bs{X};\bs{Y}|\bs{H})=NI(\bs{x};\bs{y})+L\mathcal{D}_{\text{KL}}(\mathcal{N}_c(0,\eta)||\mathcal{N}_c(0,1))$ with $\eta=\tilde{B}\Sigma_w$ and $\mathcal{D}_{\text{KL}}$ being Kullback-Leibler (KL) divergence. That means, there exists the mutual information loss for arbitrary inputs and SNRs and the loss is identified as a divergence between two Gaussian distributions, which is firstly conjectured by M{\"u}ller \cite{Muller2002channel}.
\end{remark}

\section{Replica Analysis}
\label{Sec:RM}
\subsection{Evaluation of Free Energy}
In this section, we apply the replica method derived from statistic physics to calculate the following free energy which is also the key term of mutual information in (\ref{Equ:mutual})
\begin{align}
\mathcal{F}=-\lim_{N\uparrow \infty}\frac{1}{N}\mathbb{E}_{\bs{Y},\bs{H}}\left\{\log \mathcal{P}(\bs{Y}|\bs{H})\right\}.
\end{align}
The evaluation of free energy above is extremely difficult due to the expectation-log operation. To facilitate the calculation of free energy, $\mathcal{F}$ can be rewritten as\footnote{
The following formula is applied from right to left
$$
\lim_{\tau\downarrow 0}\frac{\partial }{\partial \tau}\log \mathbb{E}\{\Theta^{\tau}\}=\lim_{\tau\downarrow 0}\frac{\mathbb{E}\{\Theta^{\tau}\log \Theta\}}{\mathbb{E}\{\Theta^{\tau}\}}=\mathbb{E}\{\log \Theta \},
$$
where $\Theta$ is any positive random variable.
}
\begin{align}
\mathcal{F}=-\lim_{N\uparrow \infty}\frac{1}{N}\lim_{\tau\downarrow 0}\frac{\partial }{\partial \tau}\log \mathbb{E}\left\{\mathcal{P}^{\tau}(\bs{Y}|\bs{H})\right\},
\label{Equ:Free}
\end{align}
where $\mathcal{P}(\bs{Y}|\bs{H})$ is the partition function with the form of
\begin{align}
\nonumber
\mathcal{P}(\bs{Y}|\bs{H})
&=\int \mathcal{P}(\bs{Y}|\bs{H},\bs{X})\mathcal{P}(\bs{X})\text{d}\bs{X}\\
&=\int \mathcal{P}(\bs{Y}|\bs{Z})\delta\left(\bs{Z}-\bs{HX}\right)\text{d}\bs{Z}\mathcal{P}(\bs{X})\text{d}\bs{X}.
\end{align}
The general operation for (\ref{Equ:Free}) assumes that $\lim_{N\uparrow \infty}$ and $\lim_{\tau\downarrow 0}$ can be exchanged. We first calculate
\begin{align}
\nonumber
\!\!\!\!\mathbb{E}\left\{\mathcal{P}^{\tau}(\bs{Y}|\bs{H})\right\}
&=\int_{\bs{Y},\bs{H}}\prod_{a=0}^{\tau}\int_{\bs{X}^{(a)}}\mathcal{P}(\bs{Y}|\bs{H},\bs{X}^{(a)})\\
\nonumber
&\quad \times \mathcal{P}(\bs{X}^{(a)})\text{d}\bs{X}^{(a)}\mathcal{P}(\bs{H})\text{d}\bs{H}\text{d}\bs{Y}\\
\nonumber
&=\int \mathcal{P}(\bs{Y}|\bs{H},\boldsymbol{\mathcal{X}})\mathcal{P}(\boldsymbol{\mathcal{X}})\mathcal{P}(\bs{H})\text{d}\bs{H}\text{d}\boldsymbol{\mathcal{X}}\text{d}\bs{Y}\\
&=\mathbb{E}_{\boldsymbol{\mathcal{X}}}\left\{
\int \mathcal{P}(\bs{Y}|\boldsymbol{\mathcal{Z}})\mathbb{E}_{\bs{H}}\left\{\delta(\boldsymbol{\mathcal{Z}}-\bs{H}\boldsymbol{\mathcal{X}})\right\}\text{d}\boldsymbol{\mathcal{Z}}\text{d}\bs{Y}
\right\},
\label{Equ:RM1}
\end{align}
where the superscript $(\cdot)^{(0)}$ denotes original signal and $(\cdot)^{(a)}$ denotes replica. In addition, following definitions: $\boldsymbol{\mathcal{X}}=\{\bs{X}^{(a)}, \forall a\}$, $\mathcal{P}(\boldsymbol{\mathcal{X}})=\prod_{a=0}^{\tau}\mathcal{P}(\bs{X}^{(a)})$, $\mathcal{P}(\bs{Y}|\bs{H},\boldsymbol{\mathcal{X}})=\prod_{a=0}^{\tau}\mathcal{P}(\bs{Y}|\bs{H},\bs{X}^{(a)})$, $\boldsymbol{\mathcal{Z}}=\{\bs{Z}^{(a)},\forall a\}$, and $\mathcal{P}(\bs{Y}|\boldsymbol{\mathcal{Z}})=\prod_{a=0}^{\tau}\mathcal{P}(\bs{Y}|\bs{Z}^{(a)})$ are applied. To deal with the inner expectation operation in (\ref{Equ:RM1}), applying PDF-to-RV lemma\footnote{
Let ${\boldsymbol{w}}\in \mathbb{C}^p$ and $\boldsymbol{u} \in \mathbb{C}^q$ be two RVs, and $g:\mathbb{C}^p\rightarrow \mathbb{C}^q$ be a generic mapping. Then, $\boldsymbol{u}=g({\boldsymbol{w}})$ if and only if the PDF $\mathcal{P}(\boldsymbol{u})\propto \int \delta(\boldsymbol{u}-g(\boldsymbol{w}))p_{{\boldsymbol{w}}}(\boldsymbol{w})\text{d}\boldsymbol{w}$.
} \cite{pfister2014compressed} and central limit theorem (CLT), each block $\bs{z}_{\ell}^{(a)}$ in $\boldsymbol{\mathcal{Z}}$ tends to be a random Gaussian vector with zero mean and covariance matrix
\begin{align}
\nonumber
\mathbb{E}_{\bs{H}}\left\{\bs{z}_{\ell}^{(a)}(\bs{z}_{\ell}^{(b)})^{\dag}\right\}
&=\sum_{n=1}^N \mathbb{E}_{\bs{H}}\left\{|h_{\ell n}|^2\bs{x}_n^{(a)}(\bs{x}_n^{(b)})^{\dag}\right\}\\
&=\frac{1}{L}\sum_{n=1}^N \bs{x}_n^{(a)}(\bs{x}_n^{(b)})^{\dag}.
\label{Equ:RM2}
\end{align}

To average over $\mathcal{P}(\boldsymbol{\mathcal{X}})$, we introduce a $M\tau\times M\tau$ auxiliary matrix $\bs{Q}$ defined as
\begin{align}
1=\int \prod_{0\leq a\leq b}\delta\left(N[\bs{Q}]_{ab}-\sum_{n=1}^N\bs{x}_n^{(a)}(\bs{x}_n^{(b)})^{\dag}\right),
\label{Equ:RM3}
\end{align}
where $[\bs{Q}]_{ab}$ is $M\times M$ matrix and $(a,b)$-th block of $\bs{Q}$.

Let's define $\tilde{\bs{z}}_{\ell}=[(\bs{z}_{\ell}^{(0)})^{\text{T}}, \cdots, (\bs{z}_{\ell}^{(\tau)})^{\text{T}}]^{\text{T}}$, substituting (\ref{Equ:RM2}) and (\ref{Equ:RM3}) into (\ref{Equ:RM1}) yields
\begin{align}
\nonumber
&\frac{1}{N}\log\mathbb{E}\left\{\mathcal{P}^{\tau}(\bs{Y}|\bs{H})\right\}\\
\nonumber
&=\frac{1}{N}\log \mathbb{E}_{\bs{Q}}\left\{\int \mathcal{P}(\bs{Y}|\boldsymbol{\mathcal{Z}})\prod_{\ell=1}^L \mathcal{N}_c(\tilde{\bs{z}}_{\ell}|\bs{0},\frac{1}{\alpha}\bs{Q})\text{d}\boldsymbol{\mathcal{Z}}\text{d}\bs{Y}\right\}\\
&=\frac{1}{N}\log \mathbb{E}_{\bs{Q}}\left\{\left(\int \mathcal{P}(\bs{y}|\tilde{\bs{z}})\mathcal{N}_c(\tilde{\bs{z}}|\bs{0}\frac{1}{\alpha}\bs{Q})\text{d}\tilde{\bs{z}}\text{d}\bs{y}\right)^{L}\right\}.
\end{align}
From (\ref{Equ:RM3}), it is observed that each block of $\bs{Q}$ is the sum of $N$ weighted random vectors. By large derivation theory (LDT) \cite{touchette2011basic} , $\mathcal{P}(\bs{Q})$ can be approximated by $\mathcal{P}(\bs{Q})\approx \exp \left(-N\mathcal{R}^{(\tau)}(\bs{Q})\right)$ with rate function
\begin{align}
\!\!\!\!\!\mathcal{R}^{(\tau)}(\bs{Q})=\sup_{\tilde{\bs{Q}}}\left\{\text{Tr}(\tilde{\bs{Q}}\bs{Q})-\log \mathbb{E}_{\tilde{\bs{x}}}\left\{\exp \left(\tilde{\bs{x}}^{\dag}\tilde{\bs{Q}}\tilde{\bs{x}}\right)\right\}\right\},
\end{align}
where $\tilde{\bs{x}}$ is $M\tau$ column vector with the form of $\tilde{\bs{x}}=[(\bs{x}^{(0)})^{\text{T}}, \cdots, (\bs{x}^{(\tau)})^{\text{T}}]^{\text{T}}$. Further, by Varahan's theorem \cite{touchette2009large}
\begin{align}
\nonumber
\!\!\!\frac{1}{N}\log \mathbb{E}\left\{\mathcal{P}^{\tau}(\bs{Y}|\bs{H})\right\}
&=\sup_{\bs{Q}}\left\{\alpha G^{(\tau)}(\bs{Q})-\mathcal{R}^{\tau}(\bs{Q})\right\}\\
\nonumber
&=\sup_{\bs{Q}}\inf_{\tilde{\bs{Q}}}\left\{
\alpha G^{(\tau)}(\bs{Q})-\text{Tr}(\tilde{\bs{Q}}\bs{Q})\right.\\
\nonumber
&\quad +\left.\log \mathbb{E}_{\tilde{\bs{x}}}\left\{\exp \left(\tilde{\bs{x}}^{\dag}\tilde{\bs{Q}}\tilde{\bs{x}}\right)\right\}
\right\}\\
&=\sup_{\bs{Q}}\inf_{\tilde{\bs{Q}}}T(\bs{Q},\tilde{\bs{Q}}),
\label{Equ:T}
\end{align}
where $G^{(\tau)}(\bs{Q})=\log \int \prod_{a=0}^{\tau}\mathcal{P}(\bs{y}|\bs{z}^{(a)})\mathcal{N}_c(\tilde{\bs{z}}|\bs{0},\frac{1}{\alpha}\bs{Q})\text{d}\tilde{\bs{z}}\text{d}\bs{y}$.

Differentiating $T(\bs{Q}, \tilde{\bs{Q}})$ w.r.t. $\bs{Q}$ and $\tilde{\bs{Q}}$ yields the saddle point equations which satisfy
\begin{subequations}
\begin{align}
\tilde{\bs{Q}}&=-\alpha [\bs{Q}^{-1}-\alpha \bs{Q}^{-1}\mathbb{E}\{\tilde{\bs{z}}\tilde{\bs{z}}^{\dag}\}\bs{Q}^{-1}],
\label{Equ:SD1}\\
\bs{Q}&=\frac{\mathbb{E}\left\{\tilde{\bs{x}}\tilde{\bs{x}}^{\dag}\exp \left(\tilde{\bs{x}}^{\dag}\tilde{\bs{Q}}\tilde{\bs{x}}\right)\right\}}{\mathbb{E}\left\{\exp \left(\tilde{\bs{x}}^{\dag}\tilde{\bs{Q}}\tilde{\bs{x}}\right)\right\}},
\label{Equ:SD2}
\end{align}
\label{Equ:SD}
\end{subequations}
where the expectation in (\ref{Equ:SD1})  is taken over
\begin{align}
\mathcal{P}(\tilde{\bs{z}})=\frac{\int \prod_{a=0}^{\tau}\mathcal{P}(\bs{y}|\bs{z}^{(a)})\mathcal{N}_c(\tilde{\bs{z}}|\bs{0},\frac{1}{\alpha}\bs{Q})\text{d}\bs{y}}
{\int \prod_{a=0}^{\tau}\mathcal{P}(\bs{y}|\bs{z}^{(a)})\mathcal{N}_c(\tilde{\bs{z}}|\bs{0},\frac{1}{\alpha}\bs{Q})\text{d}\bs{z}\text{d}\bs{y}}.
\end{align}
To derive (\ref{Equ:SD1}), the following note\footnote{
Given $\bs{x}\in \mathbb{C}^N$ following $\mathcal{N}_c(\bs{x}|\bs{0}, \alpha^{-1}\bs{Q})$, the partial derivation of $\mathcal{N}_c(\bs{x}|\bs{0}, \alpha^{-1}\bs{Q})$ w.r.t. $\bs{Q}$ is $\frac{\partial \mathcal{N}_c(\bs{x}|\bs{0}, \alpha^{-1}\bs{Q})}{\partial \bs{Q}}
=-(\bs{Q}^{-1}-\alpha \bs{Q}^{-1}\bs{xx}^{\dag}\bs{Q}^{-1})\mathcal{N}_c(\bs{x}|\bs{0}, \alpha^{-1}\bs{Q})$.
} is useful.

We here outline the procedures of calculating the free energy. The point $(\bs{Q}, \tilde{\bs{Q}})$  satisfying the saddle point equations (\ref{Equ:SD}) achieves the extremal condition in (\ref{Equ:T}) for general $\tau$ and let it be $(\bs{Q}^{\star}, \tilde{\bs{Q}}^{\star})$. Taking the partial derivation of $T(\bs{Q}^{\star}, \tilde{\bs{Q}}^{\star})$ w.r.t. $\tau$ and letting $\tau\rightarrow 0$ yields minus free energy.
Since $T(\bs{Q}, \tilde{\bs{Q}})$ is differentiable that admits a minimum and a maximum, following the result of \cite[Appendix G]{tulino2013support}, we first determine $(\bs{Q}^{\star}(\tau), \tilde{\bs{Q}}^{\star}(\tau))|_{\tau\downarrow 0}$ that satisfies the saddle pint equations (\ref{Equ:SD}). Then, replacing it in (\ref{Equ:T}) and differentiating the resulting expression w.r.t. $\tau$ and letting $\tau\rightarrow 0$ yields the same result as $\frac{\partial }{\partial \tau }T(\bs{Q}^{\star}, \tilde{\bs{Q}}^{\star})|_{\tau\downarrow 0}$.

In fact, solving the saddle point equations (\ref{Equ:SD1})-(\ref{Equ:SD2}) is prohibitive except the simplest case such as both prior and likelihood being Gaussian. For facilitating evaluation, we restrict that both $\bs{Q}^{\star}$ and $\tilde{\bs{Q}}^{\star}$  follow replica symmetric structure.

\begin{assumption}[Replica symmetry]
It is assumed that both $\bs{Q}^{\star}$ and $\tilde{\bs{Q}}^{\star}$ are replica symmetric\footnote{
In \cite{tulino2013support}, at the beginning the entries of auxiliary matrices are set as complex number, but they are verified to be real-valued after some algebras. Inspired by this work, we set each block of $\bs{Q}^{\star}$ and $\tilde{\bs{Q}}^{\star}$ is real-valued symmetric matrix.
}
\begin{subequations}
\begin{align}
\bs{Q}^{\star}&=\bs{I}\otimes (\bs{A}-\bs{B})+\bs{11}^{\rm{T}}\otimes \bs{B},\\
\tilde{\bs{Q}}^{\star}&=\bs{I}\otimes (\tilde{\bs{A}}-\tilde{\bs{B}})+\bs{11}^{\rm{T}}\otimes \tilde{\bs{B}},
\end{align}
\end{subequations}
where $\bs{1}$ is $(\tau+1)\times 1$ column vector with its all elements being $1$, $\bs{1}^{\rm{T}}$ denotes its transposition, and $\bs{11}^{\rm{T}}$ refers to $(\tau+1)\times (\tau+1)$ matrix whose elements are all 1. The parameters $\bs{A}$, $\bs{B}$, $\tilde{\bs{A}}$, and $\tilde{\bs{B}}$ are all $M\times M$ real-valued positive definite matrices. Besides, it is assumed that there exists invertible matrix $\bs{P}$ such that $\bs{P}\bs{A}\bs{P}^{-1}=\bs{\Lambda}_A$, $\bs{P}\bs{B}\bs{P}^{-1}=\bs{\Lambda}_B$ with $\bs{\Lambda}_A$, $\bs{\Lambda}_B$ being diagonal matrix whose diagonal elements are non-negative.
\end{assumption}

By replica symmetry assumption, it naturally implies that $\bs{Q}_z=\mathbb{E}\{\tilde{\bs{z}}\tilde{\bs{z}}^{\dag}\}$ is also replica symmetric, i.e.,
\begin{align}
\bs{Q}_z=\bs{I}\otimes (\bs{A}_z-\bs{B}_z)+\bs{11}^{\text{T}}\otimes \bs{B}_z.
\label{Equ:Az}
\end{align}
Before giving the procedures of solving the saddle point equations (\ref{Equ:SD1})-(\ref{Equ:SD2}), we introduce the following Lemma.
\begin{lemma}
\label{Lemma:1}
Provided that $\bs{A},\bs{B}$ are two $M\times M$ symmetric matrices, $\bs{Q}$ is block symmetric with the form of $\bs{Q}=\bs{I}\otimes (\bs{A}-\bs{B})+\bs{11}^{\rm{T}}\otimes \bs{B}$, then the inverse matrix of $\bs{Q}$ is
\begin{align}
\nonumber
\bs{Q}^{-1}&=\bs{I}\otimes (\bs{A}-\bs{B})^{-1}-\bs{11}^{\rm{T}} \\
&\otimes [(\bs{A}-\bs{B})\bs{B}^{-1}(\bs{A}-\bs{B})+(\tau+1)(\bs{A}-\bs{B})]^{-1}.
\end{align}
\proof See Appendix A.
\end{lemma}

By Lemma \ref{Lemma:1}, let's define $\bs{E}=\alpha(\bs{A}-\bs{B})^{-1}, \bs{F}=\alpha [(\bs{A}-\bs{B})\bs{B}^{-1}(\bs{A}-\bs{B})+(\bs{A}-\bs{B})]^{-1}$ to represent
\begin{align}
\alpha(\bs{Q}^{\star})^{-1}|_{\tau\downarrow 0}=\bs{I}\otimes \bs{E}-\bs{11}^{\text{T}}\otimes \bs{F}.
\end{align}
With the definitions above, we have
\begin{align}
\nonumber
&\exp \left(-\alpha \tilde{\bs{z}}^{\dag}(\bs{Q}^{\star})^{-1}\tilde{\bs{z}}\right)\\
\nonumber
&=\exp \left[-\sum_{a=0}^{\tau}(\bs{z}^{(a)})^{\dag}(\bs{E}-\bs{F})\bs{z}^{(a)}+2\sum_{0\leq a<b}\Re\{(\bs{z}^{(a)})^{\dag}\bs{F}\bs{z}^{(b)}\}\right]\\
&=\exp \left[-\sum_{a=0}^{\tau}(\bs{z}^{(a)})^{\dag}\bs{E}\bs{z}^{(a)}+\left\|\bs{F}^{\frac{1}{2}}\sum_{a=0}^{\tau}\bs{z}^{(a)}\right\|^2\right].
\label{Equ:RM6}
\end{align}
It is worth noting that due to the replica symmetric assumption, it implies that $\bs{F}$ is positive definite matrix. Thus, $\bs{F}^{\frac{1}{2}}$ exists and is also positive definite matrix.

We use the complex circularly symmetric version of the vector Hubbard-Stratonovich transform \cite{hubbard1959calculation}:
\begin{align}
\!\!\!e^{\|\bs{x}\|^2}=(\eta/\pi)^M \int \exp \left(-\eta \|\boldsymbol{\xi}\|^2+2\sqrt{\eta}\Re\left\{\bs{x}^{\dag}\boldsymbol{\xi}\right\}\right)\text{d}\boldsymbol{\xi},
\end{align}
where $\bs{x},\boldsymbol{\xi}\in \mathbb{C}^M$ and $\eta\in \mathbb{R}_{+}$. From (\ref{Equ:RM6}), we obtain
\begin{align}
\nonumber
(\ref{Equ:RM6})
&=\int (\eta/\pi)^M \exp \left[
-\eta\|\boldsymbol{\xi}\|^2-\sum_{a=0}^{\tau}(\bs{z}^{(a)})^{\dag}\bs{E}\bs{z}^{(a)}\right.\\
&\left.\quad +2\sqrt{\eta}\Re\left\{\left(\bs{F}^{\frac{1}{2}}\sum_{a=0}^{\tau}\bs{z}^{(a)}\right)^{\dag}\boldsymbol{\xi}\right\}
\right]\text{d}\boldsymbol{\xi}.
\end{align}
Using this decoupling result, we calculate the denominator of $\bs{A}_z$ in (\ref{Equ:Az}) as below
\begin{align}
\nonumber
&\int \prod_{a=0}^{\tau}\mathcal{P}(\bs{y}|\bs{z}^{(a)})\mathcal{N}_c(\tilde{\bs{z}}|\bs{0},\frac{1}{\alpha}\bs{Q}^{\star})\text{d}\tilde{\bs{z}}\text{d}\bs{y}\\
\nonumber
&=C\int_{\bs{y},\boldsymbol{\xi}} \left[\int_{\bs{z}} \mathcal{P}(\bs{y}|\bs{z})\exp \left[-\bs{z}^{\dag}\bs{E}\bs{z} \right.\right.\\
\nonumber
&\quad \left.\left.\left.+2\sqrt{\eta}\Re\left\{\left(\bs{F}^{\frac{1}{2}}\bs{z}\right)^{\dag}\boldsymbol{\xi}\right\}\right]\text{d}\bs{z}\right]^{\tau+1}\right|_{\tau\downarrow 0}\\
\nonumber
&\quad \times \left(\frac{\eta}{\pi}\right)^M\exp \left(-\eta\|\boldsymbol{\xi}\|^2\right)\text{d}\boldsymbol{\xi}\text{d}\bs{y}\\
\nonumber
&=C\int \mathcal{P}(\bs{y}|\bs{z})\det(\pi\bs{E}^{-1})\mathcal{N}_c(\bs{z}|\sqrt{\eta}\bs{E}^{-1}\bs{F}^{\frac{1}{2}}\boldsymbol{\xi},\bs{E}^{-1})\\
\nonumber
&\quad \times \left(\frac{\eta}{\pi}\right)^M\exp \left(-\eta\boldsymbol{\xi}^{\dag}(\bs{I}-\bs{F}^{\frac{1}{2}}\bs{E}^{-1}\bs{F}^{\frac{1}{2}})\boldsymbol{\xi}\right)\text{d}\bs{z}\text{d}\boldsymbol{\xi}\text{d}\bs{y}\\
&=C \pi^M\det(\bs{E}(\bs{I}-\bs{F}^{\frac{1}{2}}\bs{E}^{-1}\bs{F}^{\frac{1}{2}}))^{-1},
\label{Equ:Azden}
\end{align}
where $C=\det(\pi \alpha^{-1}\bs{Q}^{\star})^{-1}$ is a constant and will be eliminated.

The numerator of $\bs{A}_z$ in (\ref{Equ:Az}) is
\begin{align}
\nonumber
&\int \bs{z}^{(0)}(\bs{z}^{(0)})^{\dag}\prod_{a=0}^{\tau}\mathcal{P}(\bs{y}|\bs{z}^{(a)})\mathcal{N}_c(\tilde{\bs{z}}|\bs{0},\frac{1}{\alpha}\bs{Q}^{\star})\text{d}\bs{z}\text{d}\bs{y}\\
\nonumber
&=C\int \bs{z}\bs{z}^{\dag}\det(\pi\bs{E}^{-1})\mathcal{N}_c(\bs{z}|\sqrt{\eta}\bs{E}^{-1}\bs{F}^{\frac{1}{2}}\boldsymbol{\xi},\bs{E}^{-1})\\
\nonumber
&\qquad \times (\eta/\pi)^M\mathcal{N}_c(\boldsymbol{\xi}|\bs{0}, \eta^{-1}(\bs{I}-\bs{F}^{\frac{1}{2}}\bs{E}^{-1}\bs{F}^{\frac{1}{2}})^{-1})\\
\nonumber
&\qquad \times \det(\pi \eta^{-1}(\bs{I}-\bs{F}^{\frac{1}{2}}\bs{E}^{-1}\bs{F}^{\frac{1}{2}})^{-1})\text{d}\bs{z}\text{d}\boldsymbol{\xi}\\
\nonumber
&=C\pi^M\det(\bs{E}(\bs{I}-\bs{F}^{\frac{1}{2}}\bs{E}^{-1}\bs{F}^{\frac{1}{2}}))^{-1}\\
\nonumber
&\qquad \times (\bs{E}^{-1}+\bs{E}^{-1}\bs{F}^{\frac{1}{2}}(\bs{I}-\bs{F}^{\frac{1}{2}}\bs{E}^{-1}\bs{F}^{\frac{1}{2}})^{-1}\bs{F}^{\frac{1}{2}}\bs{E}^{-1})\\
\nonumber
&=C\pi^M\det(\bs{E}(\bs{I}-\bs{F}^{\frac{1}{2}}\bs{E}^{-1}\bs{F}^{\frac{1}{2}}))^{-1}\\
&\qquad \times (\bs{E}^{-1}+(\bs{E}\bs{F}^{-1}\bs{E}-\bs{E})^{-1}).
\label{Equ:Aznum}
\end{align}

Combining (\ref{Equ:Azden}) and (\ref{Equ:Aznum}) obtains
\begin{align}
\bs{A}_z&=\frac{(\ref{Equ:Aznum})}{(\ref{Equ:Azden}) }=\bs{E}^{-1}+(\bs{E}\bs{F}^{-1}\bs{E}-\bs{E})^{-1}=\alpha^{-1}\bs{A}.
\end{align}

On the other hand, the numerator of $\bs{B}_z$ is calculated in (\ref{Equ:C1}).
\begin{figure*}
\setcounter{equation}{42}
\begin{align}
\nonumber
&\int \bs{z}^{(0)}(\bs{z}^{(1)})^{\dag}\prod_{a=0}^{\tau}\mathcal{P}(\bs{y}|\bs{z}^{(a)})\mathcal{N}_c(\tilde{\bs{z}}|\bs{0},\frac{1}{\alpha}\bs{Q}^{\star})\text{d}\bs{z}\text{d}\bs{y}\\
\nonumber
&=C\int_{\bs{y},\boldsymbol{\xi}}  \int_{\bs{z}^{(0)}}\bs{z}^{(0)}\mathcal{P}(\bs{y}|\bs{z}^{(0)})
\exp \left[-(\bs{z}^{(0)})^{\dag}\bs{E}\bs{z}^{(0)}+2\sqrt{\eta}\Re\left\{\left(\bs{F}^{\frac{1}{2}}\bs{z}^{(0)}\right)^{\dag}\boldsymbol{\xi}\right\}\right]\text{d}\bs{z}^{(0)}\\
\nonumber
&\qquad \times \int_{\bs{z}^{(1)}}(\bs{z}^{(1)})^{\dag}\mathcal{P}(\bs{y}|\bs{z}^{(1)})
\exp \left[-(\bs{z}^{(1)})^{\dag}\bs{E}\bs{z}^{(1)}+2\sqrt{\eta}\Re\left\{\left(\bs{F}^{\frac{1}{2}}\bs{z}^{(1)}\right)^{\dag}\boldsymbol{\xi}\right\}\right]\text{d}\bs{z}^{(1)}\\
\nonumber
&\qquad \times  \left.
\left[\int \mathcal{P}(\bs{y}|\bs{z})\exp \left[-\bs{z}^{\dag}\bs{E}\bs{z}+2\sqrt{\eta}\Re\left\{\left(\bs{F}^{\frac{1}{2}}\bs{z}\right)^{\dag}\boldsymbol{\xi}\right\}\right]\text{d}\bs{z}\right]^{\tau-1}
\right|_{\tau\downarrow 0}\left(\frac{\eta}{\pi}\right)^M\exp \left(-\eta\|\boldsymbol{\xi}\|^2\right)\text{d}\boldsymbol{\xi}\text{d}\bs{y}\\
\nonumber
&=C\int \frac{\left[\int\bs{z}\mathcal{P}(\bs{y}|\bs{z})\mathcal{N}_c(\bs{z}|\sqrt{\eta}\bs{E}^{-1}\bs{F}^{\frac{1}{2}}\boldsymbol{\xi},\bs{E}^{-1})\text{d}\bs{z}\right]
\left[\int\bs{z}\mathcal{P}(\bs{y}|\bs{z})\mathcal{N}_c(\bs{z}|\sqrt{\eta}\bs{E}^{-1}\bs{F}^{\frac{1}{2}}\boldsymbol{\xi},\bs{E}^{-1})\text{d}\bs{z}\right]^{\dag}
}{\int \mathcal{P}(\bs{y}|\bs{z})\mathcal{N}_c(\bs{z}|\sqrt{\eta}\bs{E}^{-1}\bs{F}^{\frac{1}{2}}\boldsymbol{\xi},\bs{E}^{-1})\text{d}\bs{z}}\\
&\qquad \times \pi^M \det(\bs{E}(\bs{I}-\bs{F}^{\frac{1}{2}}\bs{E}^{-1}\bs{F}^{-\frac{1}{2}}))^{-1}\mathcal{N}_c(\boldsymbol{\xi}|\bs{0}, \eta^{-1}(\bs{I}-\bs{F}^{\frac{1}{2}}\bs{E}^{-1}\bs{F}^{\frac{1}{2}})^{-1})\text{d}\boldsymbol{\xi}\text{d}\bs{y}.
\label{Equ:C1}
\end{align}
\hrulefill
\end{figure*}
Let's define $\boldsymbol{\zeta}=\sqrt{\eta}(\bs{I}-\bs{F}^{\frac{1}{2}}\bs{E}^{-1}\bs{F}^{\frac{1}{2}})^{\frac{1}{2}}\boldsymbol{\xi}$, then we have
$\boldsymbol{\xi}=\frac{1}{\sqrt{\eta}}(\bs{I}-\bs{F}^{\frac{1}{2}}\bs{E}^{-1}\bs{F}^{\frac{1}{2}})^{-\frac{1}{2}}\boldsymbol{\zeta}$ and $\text{d}\boldsymbol{\xi}=
\det(\frac{1}{\eta}(\bs{I}-\bs{F}^{\frac{1}{2}}\bs{E}^{-1}\bs{F}^{\frac{1}{2}})^{-1})\text{d}\boldsymbol{\zeta}$. By changing variables and replica symmetric assumption, the following can be obtained
\begin{align}
\bs{B}_z=\frac{(\ref{Equ:C1})}{(\ref{Equ:Azden})}
&=\int \frac{\bs{U}_{\bs{y}|\bs{z}}\left(\frac{1}{\sqrt{\alpha}}\bs{B}^{\frac{1}{2}}\boldsymbol{\zeta}, \frac{1}{\alpha}(\bs{A}-\bs{B})\right)}
{v_{\bs{y}|\bs{z}}\left(\frac{1}{\sqrt{\alpha}}\bs{B}^{\frac{1}{2}}\boldsymbol{\zeta}, \frac{1}{\alpha}(\bs{A}-\bs{B})\right)}\text{\rm{D}}\boldsymbol{\zeta}{\rm{d}}\bs{y},
\label{Equ:C2}
\end{align}
where $\text{D}\boldsymbol{\zeta}=\mathcal{N}_c(\boldsymbol{\zeta}|\bs{0},\bs{I})\text{d}\boldsymbol{\xi}$ is complex-valued vector Gaussian measure and the function $\bs{U}_{\bs{y}|\bs{z}}(\cdot)$ and $v_{\bs{y}|\bs{z}}(\cdot)$ are defined in (\ref{Equ:def}).

Based on the fact $\bs{A}_z=\frac{1}{\alpha}\bs{A}$, solving the saddle point equation (\ref{Equ:SD1}) yields
\begin{align}
\tilde{\bs{A}}&=\bs{0},\\
\tilde{\bs{B}}&=(\bs{A}-\bs{B})^{-1}(\alpha^2 \bs{B}_z-\alpha\bs{B})(\bs{A}-\bs{B})^{-1}.
\label{Equ:counter}
\end{align}
The details of obtaining $\tilde{\bs{A}}$ and $\tilde{\bs{B}}$ is presented in Appendix B.

Based on replica symmetric assumption, the term in (\ref{Equ:SD2}) can be expressed as
\begin{align}
\nonumber
\mathbb{E}\left\{\exp \left(\bs{x}^{\dag}\tilde{\bs{Q}}^{\star}\bs{x}\right)\right\}
&=\mathbb{E}\left\{
\exp \left(\sum_{a=0}^{\tau}(\bs{x}^{(a)})^{\dag}\tilde{\bs{A}}\bs{x}^{(a)}\right.\right.\\
\nonumber
&\left.\left.+2\sum_{0\leq a<b}\Re\left\{(\bs{x}^{(a)})^{\dag}\tilde{\bs{B}}\bs{x}^{(b)}\right\}\right)
\right\}\\
\nonumber
&=\mathbb{E}\left\{
\exp \left(\left\|\sum_{a=0}^{\tau}\tilde{\bs{B}}^{\frac{1}{2}}\bs{x}^{(a)}\right\|^2\right.\right.\\
&\left.\left.+\sum_{a=0}^{\tau}(\bs{x}^{(a)})^{\dag}(\tilde{\bs{A}}-\tilde{\bs{B}})\bs{x}^{(a)}\right)
\right\}.
\label{Equ:C3}
\end{align}
Also by complex circularly symmetric version of the vector Hubbard-Stratonovich transform and the fact $\tilde{\bs{A}}=\bs{0}$
\begin{align}
\nonumber
\!\!\!\!\!\!(\ref{Equ:C3})
&=\mathbb{E}_{\bs{x}}\left\{\int_{\boldsymbol{\xi}} (\eta/\pi)^M
\exp \left(
-\eta\|\boldsymbol{\xi}\|^2-\sum_{a=0}^{\tau}(\bs{x}^{(a)})^{\dag}\tilde{\bs{B}}\bs{x}^{(a)}\right.\right.\\
\nonumber
&\quad \left.\left.+2\sqrt{\eta}\Re\left\{\left(\tilde{\bs{B}}^{\frac{1}{2}}\sum_{a=0}^{\tau}\bs{x}^{(a)}\right)^{\dag}\boldsymbol{\xi}\right\}
\right)\text{d}\boldsymbol{\xi}
\right\}\\
\nonumber
&=\int (\eta/\pi)^M \left[\int_{\bs{x}}
\exp \left(-\eta\|\boldsymbol{\xi}\|^2+2\sqrt{\eta}\Re\left\{\left(\tilde{\bs{B}}^{\frac{1}{2}}\bs{x}\right)^{\dag}\boldsymbol{\xi}\right\}\right.\right.\\
\nonumber
&\quad \left.\left.\left.-\bs{x}^{\dag}\tilde{\bs{B}}\bs{x}
\right)\mathcal{P}_{\textsf{X}}(\bs{x})
\right]^{\tau+1}\exp \left(\eta\tau\|\boldsymbol{\xi}\|^2\right)\right|_{\tau\downarrow 0}\text{d}\boldsymbol{\xi}\\
&=\int
(\frac{\eta}{\pi})^M\exp \left[-\eta\|\boldsymbol{\xi}-\frac{1}{\sqrt{\eta}}\tilde{\bs{B}}^{\frac{1}{2}}\bs{x}\|^2\right]\mathcal{P}_{\textsf{X}}(\bs{x})\text{d}\bs{x}\text{d}\boldsymbol{\xi}.
\label{Equ:C4}
\end{align}
Defining $\boldsymbol{\xi}=\frac{1}{\sqrt{\eta}}\tilde{\bs{B}}^{\frac{1}{2}}\boldsymbol{\zeta}$ as well as  $\text{d}\boldsymbol{\xi}=\eta^{-M}\det(\tilde{\bs{B}})\text{d}\boldsymbol{\zeta}$, (\ref{Equ:C4}) becomes
\begin{align}
\nonumber
\mathbb{E}\left\{\bs{x}^{\dag}\tilde{\bs{Q}}^{\star}\bs{x}\right\}
&=\int \mathcal{P}_{\textsf{X}}(\bs{x})
\pi^{-M}\det(\tilde{\bs{B}})^{-1}\\
\nonumber
&\quad \times \exp \left[-(\boldsymbol{\zeta}-\bs{x})^{\dag}\tilde{\bs{B}}(\boldsymbol{\zeta}-\bs{x})\right]\text{d}\bs{x}\text{d}\boldsymbol{\zeta}\\
\nonumber
&=\int \mathcal{N}_{c}\left(\boldsymbol{\zeta}|\bs{x}, \tilde{\bs{B}}^{-1}\right)\mathcal{P}_{\textsf{X}}(\bs{x})\text{d}\bs{x}\text{d}\boldsymbol{\zeta}\\
&=1.
\end{align}
Thus
\begin{align}
\bs{A}&=\mathbb{E}\left\{\bs{x}^{(0)}(\bs{x}^{(0)})^{\dag}\exp \left(\bs{x}^{\dag}\tilde{\bs{Q}}^{\star}\bs{x}\right)\right\},\\
\bs{B}&=\mathbb{E}\left\{\bs{x}^{(0)}(\bs{x}^{(1)})^{\dag}\exp \left(\bs{x}^{\dag}\tilde{\bs{Q}}^{\star}\bs{x}\right)\right\}.
\end{align}
Below we give a brief procedures to calculate $\bs{A}$ and $\bs{B}$.
\begin{align}
\nonumber
&\bs{A}
=\int_{\boldsymbol{\xi}} (\eta/\pi)^M
\left[\int_{\bs{x}} \bs{x}^{(0)}(\bs{x}^{(0)})^{\dag}
\exp \left(-\eta\|\boldsymbol{\xi}\|^2\right.\right.\\
\nonumber
& \left.\left.+2\sqrt{\eta}\Re\left\{\left(\tilde{\bs{B}}^{\frac{1}{2}}\bs{x}^{(0)}\right)^{\dag}\boldsymbol{\xi}\right\}
-(\bs{x}^{(0)})^{\dag}\tilde{\bs{B}}\bs{x}^{(0)}
\right)\mathcal{P}_{\textsf{X}}(\bs{x}^{(0)})
\text{d}\bs{x}^{(0)}\right]\\
\nonumber
& \times\left.
\left[\int_{\bs{x}}
\exp \left(2\sqrt{\eta}\Re\left\{\left(\tilde{\bs{B}}^{\frac{1}{2}}\bs{x}\right)^{\dag}\boldsymbol{\xi}\right\}
-\bs{x}^{\dag}\tilde{\bs{B}}\bs{x}
\right)\mathcal{P}_{\textsf{X}}(\bs{x})\text{d}\bs{x}
\right]^{\tau}\right|_{\tau\downarrow 0}\text{d}\boldsymbol{\xi}\\
\nonumber
&=\int \bs{xx}^{\dag}\mathcal{N}_{c}\left(\boldsymbol{\zeta}|\bs{x}, \tilde{\bs{B}}^{-1}\right)\mathcal{P}_{\textsf{X}}(\bs{x})\text{d}\bs{x}\text{d}\boldsymbol{\zeta}\\
&=\bs{\Xi}_x.
\label{Equ:D1}
\end{align}
The parameter $\bs{B}$ is calculated in (\ref{Equ:Bexpress}).
\begin{figure*}
\setcounter{equation}{52}
\begin{align}
\nonumber
\bs{B}
&=\int_{\boldsymbol{\xi}} (\eta/\pi)^M
\left[\int_{\bs{x}} \bs{x}^{(0)}
\exp \left(-\eta\|\boldsymbol{\xi}\|^2+2\sqrt{\eta}\Re\left\{\left(\tilde{\bs{B}}^{\frac{1}{2}}\bs{x}^{(0)}\right)^{\dag}\boldsymbol{\xi}\right\}
-(\bs{x}^{(0)})^{\dag}\tilde{\bs{B}}\bs{x}^{(0)}
\right)\mathcal{P}_{\textsf{X}}(\bs{x}^{(0)})\text{d}\bs{x}^{(0)}
\right]\\
\nonumber
&\qquad \times
\left[\int_{\bs{x}}(\bs{x}^{(1)})^{\dag}
\exp \left(-\eta\|\boldsymbol{\xi}\|^2+2\sqrt{\eta}\Re\left\{\left(\tilde{\bs{B}}^{\frac{1}{2}}\bs{x}^{(1)}\right)^{\dag}\boldsymbol{\xi}\right\}
-(\bs{x}^{(1)})^{\dag}\tilde{\bs{B}}\bs{x}^{(1)}
\right)\mathcal{P}_{\textsf{X}}(\bs{x}^{(1)})\text{d}\bs{x}^{(1)}
\right]\\
\nonumber
&\qquad \times\left.
\left[\int_{\bs{x}}
\exp \left(-\eta\|\boldsymbol{\xi}\|^2+2\sqrt{\eta}\Re\left\{\left(\tilde{\bs{B}}^{\frac{1}{2}}\bs{x}\right)^{\dag}\boldsymbol{\xi}\right\}
-\bs{x}^{\dag}\tilde{\bs{B}}\bs{x}
\right)\mathcal{P}_{\textsf{X}}(\bs{x})\text{d}\bs{x}
\right]^{\tau-1}\exp \left(\eta\tau\|\boldsymbol{\xi}\|^2\right)\right|_{\tau\downarrow 0}\text{d}\boldsymbol{\xi}\\
&=\int_{\boldsymbol{\xi}} (\eta/\pi)^M \frac{\left[\int_{\bs{x}}\bs{x} \mathcal{P}_{\textsf{X}}(\bs{x})
\exp \left(-\eta\|\boldsymbol{\xi}-\frac{1}{\sqrt{\eta}}\tilde{\bs{B}}^{\frac{1}{2}}\bs{x}\|^2\right)
\text{d}\bs{x}\right]\left[\int_{\bs{x}}\bs{x}^{\dag} \mathcal{P}_{\textsf{X}}(\bs{x})
\exp \left(-\eta\|\boldsymbol{\xi}-\frac{1}{\sqrt{\eta}}\tilde{\bs{B}}^{\frac{1}{2}}\bs{x}\|^2\right)
\text{d}\bs{x}\right]}{\int_{\bs{x}} \mathcal{P}_{\textsf{X}}(\bs{x})
\exp \left(-\eta\|\boldsymbol{\xi}-\frac{1}{\sqrt{\eta}}\tilde{\bs{B}}^{\frac{1}{2}}\bs{x}\|^2\right)
\text{d}\bs{x}}\text{d}\boldsymbol{\xi}.
\label{Equ:Bexpress}
\end{align}
\hrulefill
\end{figure*}
Defining $\boldsymbol{\xi}=\frac{1}{\sqrt{\eta}}\tilde{\bs{B}}^{\frac{1}{2}}\boldsymbol{\zeta}$,  (\ref{Equ:Bexpress}) can be simplified as
\begin{align}
\bs{B}&=\int \frac{\bs{U}_{\bs{x}}(\boldsymbol{\zeta}, \tilde{\bs{B}}^{-1})}{v_{\bs{x}}(\boldsymbol{\zeta}, \tilde{\bs{B}}^{-1})}{\rm{d}}\boldsymbol{\zeta}.
\label{Equ:D2}
\end{align}
Totally, the fixed point equation is summarized as below
\begin{subequations}
\begin{align}
\bs{B}_z&=\int \frac{\bs{U}_{\bs{y}|\bs{z}}\left(\frac{1}{\sqrt{\alpha}}\bs{B}^{\frac{1}{2}}\boldsymbol{\zeta}, \frac{1}{\alpha}(\bs{A}-\bs{B})\right)}
{v_{\bs{y}|\bs{z}}\left(\frac{1}{\sqrt{\alpha}}\bs{B}^{\frac{1}{2}}\boldsymbol{\zeta}, \frac{1}{\alpha}(\bs{A}-\bs{B})\right)}\text{\rm{D}}\boldsymbol{\zeta}{\rm{d}}\bs{y},\\
\tilde{\bs{B}}&=(\bs{A}-\bs{B})^{-1}(\alpha^2\bs{B}_z-\alpha\bs{B})(\bs{A}-\bs{B})^{-1},\\
\bs{A}&=\bs{\Xi}_x\\
\bs{B}&=\int \frac{\bs{U}_{\bs{x}}(\boldsymbol{\zeta}, \tilde{\bs{B}}^{-1})}{v_{\bs{x}}(\boldsymbol{\zeta}, \tilde{\bs{B}}^{-1})}{\rm{d}}\boldsymbol{\zeta},
\label{SD4}
\end{align}
\label{SSDD}
\end{subequations}
where the definitions $\bs{U}_{\bs{y}|\bs{z}}$ and $v_{\bs{y}|\bs{z}}$ in (\ref{Equ:def}) are applied.

Finally, we define a single-vector channel according to (\ref{Equ:D1}) and (\ref{Equ:D2}).
\begin{align}
\bs{y}=\bs{x}+\bs{n},
\label{Equ:D4}
\end{align}
where $\bs{x}\sim \mathcal{P}_{\textsf{X}}(\bs{x})$ with zero mean and $\bs{\Xi}_x$ covariance matrix,  and $\bs{n}$ follows complex Gaussian vector distribution with the form of $\mathcal{P}(\bs{w})=\mathcal{N}_c(\bs{w}|\bs{0}, \tilde{\bs{B}}^{-1})$. Then transition distribution of this single-vector channel is given by
\begin{align}
\!\!\!\!\!\!\mathcal{P}(\bs{y}|\bs{x};\tilde{\bs{B}})=\det(\pi \tilde{\bs{B}}^{-1})^{-1}\exp \left(-(\bs{y}-\bs{x})^{\dag}\tilde{\bs{B}}(\tilde{\bs{y}}-\bs{x})\right).
\end{align}
By Bayes rules, it is easy to get the posterior distribution
\begin{align}
\mathcal{P}(\bs{x}|\bs{y})
&=\frac{\mathcal{P}_{\textsf{X}}(\bs{x})\mathcal{P}(\bs{y}|\bs{x};\tilde{\bs{B}})}{\int \mathcal{P}_{\textsf{X}}(\bs{x})\mathcal{P}(\bs{y}|\bs{x};\tilde{\bs{B}})\text{d}\bs{x}}.
\end{align}
The posterior mean of $\bs{x}$ given $\bs{y}$ and its second moment are
\begin{align}
\mathbb{E}_{\bs{x}|\bs{y}}\{\bs{x}\}&=\int \bs{x}\mathcal{P}(\bs{x}|\bs{y})\text{d}\bs{x},\\
\mathbb{E}_{\bs{x}|\bs{y}}\{\bs{x}\bs{x}^{\dag}\}&=\int \bs{x}\bs{x}^{\dag}\mathcal{P}(\bs{x}|\bs{y})\text{d}\bs{x}.
\end{align}
The MMSE matrix is given by
\begin{align}
\textsf{MMSE}_{\bs{x}}
=\left\{\mathbb{E}_{\bs{x}}\{\bs{x}\bs{x}^{\dag}\}\right\}-\mathbb{E}_{\bs{y}}\left\{\mathbb{E}_{\bs{x}|\bs{y}}\{\bs{x}\}\mathbb{E}_{\bs{x}|\bs{y}}\{\bs{x}\}^{\dag}\right\},
\label{Equ:D3}
\end{align}
where $\mathcal{P}(\bs{y})=\int \mathcal{P}_{\textsf{X}}(\bs{x})\mathcal{P}(\bs{y}|\bs{x})\text{d}\bs{x}$. At this point, it is easy to verify the first term in (\ref{Equ:D3}) shares the same expression with $\bs{A}$ in (\ref{SSDD}) and the second term in (\ref{Equ:D3}) is the same as $\bs{B}$ in (\ref{SSDD}), i.e.,
\begin{align}
\bs{A}&=\mathbb{E}_{\bs{y}}\left\{\mathbb{E}_{\bs{x}|\bs{y}}\{\bs{x}\bs{x}^{\dag}\}\right\},\\
\bs{B}&=\mathbb{E}_{\bs{y}}\left\{\mathbb{E}_{\bs{x}|\bs{y}}\{\bs{x}\}\mathbb{E}_{\bs{x}|\bs{y}}\{\bs{x}\}^{\dag}\right\}.
\end{align}
Obviously, the MMSE matrix of this single-vector channel is addressed as $\textsf{MMSE}_{\bs{x}}=\bs{A}-\bs{B}$. Accordingly, the MSE associated with MMSE matrix is defined as
\begin{align}
\textsf{MSE}_{\bs{x}}=\frac{1}{M}\text{Tr}(\textsf{MMSE}_{\bs{x}})=\frac{1}{M}\text{Tr}\left(\bs{A}-\bs{B}\right).
\end{align}

We now move to calculate the free energy with the form of
\begin{align}
\nonumber
\mathcal{F}
&=-\frac{\partial }{\partial \tau}T(\bs{Q}^{\star}, \tilde{\bs{Q}}^{\star})|_{\tau\downarrow 0}\\
\nonumber
&=-\alpha \left.\frac{\partial }{\partial \tau}G^{(\tau)}(\bs{Q}^{\star})\right|_{\tau\downarrow 0}+\left.\frac{\partial }{\partial \tau}\text{Tr}(\bs{Q}^{\star}\tilde{\bs{Q}}^{\star})\right|_{\tau\downarrow 0}\\
&\quad -\left.\frac{\partial }{\partial \tau}\log \mathbb{E}_{\tilde{\bs{x}}}\left\{\exp(\tilde{\bs{x}}^{\dag}\tilde{\bs{Q}}^{\star}\tilde{\bs{x}})\right\}\right|_{\tau\downarrow 0}
\label{Equ:Freesum}.
\end{align}
Note that different from solving the saddle point equations under the precondition $\tau=0$, we here first calculate the partial derivation of $T(\bs{Q}^{\star}, \tilde{\bs{Q}}^{\star})$ w.r.t. $\tau$ in general $\tau$, i.e., $\frac{\partial }{\partial \tau}T(\bs{Q}^{\star}, \tilde{\bs{Q}}^{\star})$, and then set $\tau=0$, i.e., $\frac{\partial }{\partial \tau}T(\bs{Q}^{\star}, \tilde{\bs{Q}}^{\star})|_{\tau\downarrow 0}$. In (\ref{Equ:Freesum}), it is easy to obtain $\frac{\partial }{\partial \tau}\text{Tr}(\bs{Q}^{\star}\tilde{\bs{Q}}^{\star})|_{\tau\downarrow 0}=\text{Tr}(\bs{B}\tilde{\bs{B}})$. The difficulties lie in the terms $\frac{\partial }{\partial \tau}G^{(\tau)}(\bs{Q}^{\star})|_{\tau\downarrow 0}$ and $\frac{\partial }{\partial \tau}\log \mathbb{E}_{\tilde{\bs{x}}}\left\{\exp(\tilde{\bs{x}}^{\dag}\tilde{\bs{Q}}^{\star}\tilde{\bs{x}})\right\}|_{\tau\downarrow 0}$. For $\frac{\partial }{\partial \tau}G^{(\tau)}(\bs{Q}^{\star})|_{\tau\downarrow 0}$, applying the similar procedures as (\ref{Equ:RM6})-(\ref{Equ:Azden}) but remaining $\tau$ leads to
\begin{align}
\nonumber
&\left.\frac{\partial }{\partial \tau}G^{(\tau)}(\bs{Q}^{\star})\right|_{\tau\downarrow 0}
=\int v_{\bs{y}|\bs{z}}\left(\frac{1}{\sqrt{\alpha}}\bs{B}^{\frac{1}{2}}\boldsymbol{\zeta}, \frac{1}{\alpha}(\bs{A}-\bs{B})\right)\\
&\quad \times \log v_{\bs{y}|\bs{z}}\left(\frac{1}{\sqrt{\alpha}}\bs{B}^{\frac{1}{2}}\boldsymbol{\zeta}, \frac{1}{\alpha}(\bs{A}-\bs{B})\right)\text{D}\boldsymbol{\zeta}\text{d}\bs{y}.
\label{Equ:Free1}
\end{align}
For term $\frac{\partial }{\partial \tau}\log \mathbb{E}_{\tilde{\bs{x}}}\left\{\exp(\tilde{\bs{x}}^{\dag}\tilde{\bs{Q}}^{\star}\tilde{\bs{x}})\right\}|_{\tau\downarrow 0}$, we apply the procedures of (\ref{Equ:C3})-(\ref{Equ:C4}) but remain $\tau$, which leads to
\begin{align}
\nonumber
&\left.\frac{\partial }{\partial \tau}\log \mathbb{E}_{\tilde{\bs{x}}}\left\{\exp(\tilde{\bs{x}}^{\dag}\tilde{\bs{Q}}^{\star}\tilde{\bs{x}})\right\}\right|_{\tau\downarrow 0}
=\mathbb{E}_{\bs{y}}\{\log \mathcal{P}(\bs{y})\}+M\\
&\qquad \quad  +M\log \pi-\log \det(\tilde{\bs{B}})+\text{Tr}(\bs{A}\tilde{\bs{B}}).
\label{Equ:Free2}
\end{align}
Totally, substituting (\ref{Equ:Free1}), (\ref{Equ:Free1}) and  $\frac{\partial }{\partial \tau}\text{Tr}(\bs{Q}^{\star}\tilde{\bs{Q}}^{\star})|_{\tau\downarrow 0}=\text{Tr}(\bs{B}\tilde{\bs{B}})$ into (\ref{Equ:Freesum}) constructs the analytical result of the free energy.

\subsection{Joint Moments}
We here show the joint moments $(\bs{x}_n, \hat{\bs{x}}_n)$ of the $n$-th entry of $\bs{X}$ and corresponding MMSE estimator $\hat{\bs{X}}$ in system (\ref{Equ:system}) converge to the joint moments of $(\bs{x}, \hat{\bs{x}})$ of the equivalent single-vector channel (\ref{Equ:vecchannel}).
Following \cite{guo2005randomly}, we proceed to calculate the joint moments \cite{ghazal2000second}
\begin{align}
\mathbb{E}\{\Re(\bs{x}_n)^{\times i_{\text{R}}}\otimes \Im(\bs{x}_n)^{\times i_{\text{I}}}\otimes \Re(\hat{\bs{x}}_n)^{\times j_{\text{R}}}\otimes \Im(\hat{\bs{x}}_n)^{\times j_{\text{I}}}\},
\end{align}
for any non-negative integrals $i_{\text{R}},i_{\text{I}}, j_{\text{R}}, j_{\text{I}}$. Based on the separable posterior, we have
\begin{align}
\nonumber
&\mathbb{E}\{\Re(\bs{x}_n)^{\times i_{\text{R}}}\otimes \Im(\bs{x}_n)^{\times i_{\text{R}}}\otimes \Re(\hat{\bs{x}}_n)^{\times j_{\text{R}}}\otimes \Im(\hat{\bs{x}}_n)^{\times j_{\text{I}}} \}\\
\nonumber
&=\mathbb{E}\left\{\Re(\bs{x}_n)^{\times i_{\text{R}}}\otimes \Im(\bs{x}_n)^{\times i_{\text{R}}}
\otimes \tilde{\prod}_{u_{\text{R}}=1}^{j_{\text{R}}}\Re(\bs{x}_n^{(u_{\text{R}})})\right.\\
&\qquad \left.
\otimes \tilde{\prod}_{u_{\text{I}}=1}^{j_{\text{I}}} \Im(\bs{x}_n^{(u_{\text{I}})})
\right\},
\label{Equ:K1}
\end{align}
where $\tilde{\prod}_{u=1}^j\bs{x}_n^{(u)}=\bs{x}_n^{(1)}\otimes \cdots\otimes \bs{x}_n^{(j)}$, and  the expectation on the right hand is taken over $\prod_{u=0}^{\max\{j_{\text{R}}, \text{j}_{\text{I}}\}}\mathcal{P}(\bs{x}_{n}^{(u)},\bs{Y},\bs{H})$.

Let's define
\begin{align}
\nonumber
f(\boldsymbol{\mathcal{X}})&=\sum_{n=1}^{N'}\Re(\bs{x}_n)^{\times i_{\text{R}}}\otimes \Im(\bs{x}_n)^{\times i_{\text{I}}}\\
&\quad \otimes \tilde{\prod}_{u_{\text{R}}=1}^{j_{\text{R}}}\Re(\bs{x}_n^{(u_{\text{R}})})\otimes \tilde{\prod}_{u_{\text{I}}=1}^{j_{\text{I}}} \Im(\bs{x}_n^{(u_{\text{I}})}),
\end{align}
where $N'=\varrho N$ with $\varrho\in (0,1)$. Following \cite[Lemma 1]{guo2005randomly}
\begin{align}
\nonumber
&\frac{1}{N'}\mathbb{E}\left\{f(\boldsymbol{\mathcal{X}})\right\}=
\lim_{\tau\downarrow 0}\frac{\partial }{\partial \epsilon}\frac{1}{N'}\\
&\times \log \mathbb{E}\left\{\mathbb{E}\left\{e^{\epsilon f(\boldsymbol{\mathcal{X}})}\prod_{a=1}^{\tau}\mathcal{P}(\bs{Y}|\bs{H},\bs{X}^{(a)})\right\}\right\}|_{\epsilon=0}
\label{Equ:K3},
\end{align}
where $\boldsymbol{\mathcal{X}}=\{\bs{x}^{(a)}\}_{a=0}^{\tau}$, the inner expectation is taken over $\prod_{a=1}^{\tau}\mathcal{P}(\bs{X}^{(a)})$, and the outer expectation is taken over $\mathcal{P}(\bs{Y},\bs{H},\bs{X}^{(0)})$. It is worth noting that under $\epsilon=0$, the term $\mathbb{E}\{\mathbb{E}\{e^{\epsilon f(\boldsymbol{\mathcal{X}})}\prod_{a=1}^{\tau}\mathcal{P}(\bs{Y}|\bs{H},\bs{X}^{(a)})\}\}$ reduces to $\mathbb{E}\{\mathcal{P}^{\tau}(\bs{Y}|\bs{H})\}$ in (\ref{Equ:RM1}).

Proceeding with the same procedures as in (\ref{Equ:Free})-(\ref{Equ:T}), the following can be obtained
\begin{align}
\nonumber
&\frac{1}{N}\log \mathbb{E}\left\{\mathbb{E}\left\{e^{\epsilon f(\boldsymbol{\mathcal{X}})}\prod_{a=1}^{\tau}\mathcal{P}(\bs{Y}|\bs{H},\bs{X}^{(a)})\right\}\right\}|_{\epsilon=0}\\
& =\sup_{\bs{Q}}\inf_{\tilde{\bs{Q}}}\tilde{T}(\bs{Q},\tilde{\bs{Q}}),
\end{align}
where
\begin{align}
\nonumber
\tilde{T}(\bs{Q},\tilde{\bs{Q}})&=\sup_{\bs{Q}}\inf_{\tilde{\bs{Q}}}\left\{\alpha G^{\tau}(\bs{Q})-\text{Tr}(\tilde{\bs{Q}}\bs{Q})\right.\\
&\qquad \qquad \qquad \left. +\log \mathbb{E}\left\{e^{\epsilon \varrho g(\underline{\bs{x}})}e^{\tilde{\bs{x}}^{\dag}\tilde{\bs{Q}}\tilde{\bs{x}}}\right\}\right\},
\end{align}
where $g(\underline{\bs{x}})=\Re(\bs{x})^{\times i_{\text{R}}}\otimes \Im(\bs{x})^{\times i_{\text{R}}}\otimes \tilde{\prod}_{u_{\text{R}}=1}^{j_{\text{R}}}\Re(\bs{x}^{(u_{\text{R}})})\otimes \tilde{\prod}_{u_{\text{I}}=1}^{j_{\text{I}}} \Im(\bs{x}^{(u_{\text{I}})})$. Note that $\tilde{T}(\bs{Q},\tilde{\bs{Q}})$ shares the similar expression with $T(\bs{Q},\tilde{\bs{Q}})$ in (\ref{Equ:T}) with difference in generating moment function. Let denote $(\bs{Q}^{\star},\tilde{\bs{Q}}^{\star})$ be the saddle point of $\tilde{T}(\bs{Q},\tilde{\bs{Q}})$. Then taking the partial derivation of $\tilde{T}(\bs{Q}^{\star},\tilde{\bs{Q}}^{\star})$ w.r.t. $\epsilon$ at $\epsilon=0$ only leaves one term
\begin{align}
\nonumber
&\frac{\partial }{\partial \epsilon}\log \mathbb{E}\left\{e^{\epsilon\varrho g(\underline{\bs{x}})}e^{\tilde{\bs{x}}^{\dag}\tilde{\bs{Q}}^{\star}\bs{x}}\right\}|_{\epsilon=0}
\\
&=\varrho\frac{\mathbb{E}_{\tilde{\bs{x}}}\left\{g(\underline{\bs{x}})\exp \left(\tilde{\bs{x}}^{\dag}\tilde{\bs{Q}}^{\star}\tilde{\bs{x}}\right)\right\}}{\mathbb{E}_{\tilde{\bs{x}}}\left\{\exp \left(\tilde{\bs{x}}^{\dag}\tilde{\bs{Q}}^{\star}\tilde{\bs{x}}\right)\right\}}
\label{Equ:K4}.
\end{align}
Since $\tilde{T}(\bs{Q},\tilde{\bs{Q}})$ shares the same saddle point with $T(\bs{Q},\tilde{\bs{Q}})$ under $\epsilon=0$, similar to (\ref{Equ:C3})-(\ref{Equ:D2}) we have $\mathbb{E}\left\{\exp \left(\tilde{\bs{x}}^{\dag}\tilde{\bs{Q}}^{\star}\tilde{\bs{x}}\right)\right\}=1$ and
\begin{align}
\nonumber
&\mathbb{E}_{\tilde{\bs{x}}}\left\{g(\underline{\bs{x}})\exp \left(\tilde{\bs{x}}^{\dag}\tilde{\bs{Q}}^{\star}\tilde{\bs{x}}\right)\right\}\\
\nonumber
=&\int_{\boldsymbol{\zeta}} \int_{\bs{x}} \Re(\bs{x})^{\times i_{\text{R}}}\otimes \Im(\bs{x})^{\times i_{\text{I}}}\mathcal{P}_{\textsf{X}}(\bs{x})\mathcal{N}_c(\bs{x}|\boldsymbol{\zeta}, \tilde{\bs{B}}^{-1})\text{d}\bs{x}\\
\nonumber
&\otimes  \left[\frac{\int \Re(\bs{x})\mathcal{P}_{\textsf{X}}(\bs{x})\mathcal{N}_c(\bs{x}|\boldsymbol{\zeta}, \tilde{\bs{B}}^{-1})\text{d}\bs{x}}{\int \mathcal{P}_{\textsf{X}}(\bs{x})\mathcal{N}_c(\bs{x}|\boldsymbol{\zeta}, \tilde{\bs{B}}^{-1})\text{d}\bs{x}}\right]^{\times j_{\text{R}}}\\
\nonumber
&\otimes  \left[\frac{\int \Im(\bs{x})\mathcal{P}_{\textsf{X}}(\bs{x})\mathcal{N}_c(\bs{x}|\boldsymbol{\zeta}, \tilde{\bs{B}}^{-1})\text{d}\bs{x}}{\int \mathcal{P}_{\textsf{X}}(\bs{x})\mathcal{N}_c(\bs{x}|\boldsymbol{\zeta}, \tilde{\bs{B}}^{-1})\text{d}\bs{x}}\right]^{\times j_{\text{I}}}\text{d}\boldsymbol{\zeta}\\
=&\mathbb{E}\left\{\Re(\bs{x})^{\times i_{\text{R}}}\otimes \Im(\bs{x})^{\times i_{\text{I}}}\otimes \Re(\hat{\bs{x}})^{\times j_{\text{R}}}\otimes \Im(\hat{\bs{x}})^{\times j_{\text{I}}} \right\}.
\label{Equ:K5}
\end{align}
Combining (\ref{Equ:K1})-(\ref{Equ:K5}), we obtain
\begin{align}
\nonumber
&\frac{1}{N'}\sum_{n=1}^{N'}\mathbb{E}\left\{\Re(\bs{x}_n)^{\times i_{\text{R}}}\otimes \Im(\bs{x}_n)^{\times i_{\text{I}}}\otimes \Re(\hat{\bs{x}}_n)^{\times j_{\text{R}}}\otimes \Im(\hat{\bs{x}}_n)^{\times j_{\text{I}}}
\right\}\\
&=\mathbb{E}\left\{\Re(\bs{x})^{\times i_{\text{R}}}\otimes \Im(\bs{x})^{\times i_{\text{I}}}\otimes \Re(\hat{\bs{x}})^{\times j_{\text{R}}}\otimes \Im(\hat{\bs{x}})^{\times j_{\text{I}}} \right\},
\end{align}
where $\bs{x}$ and $\hat{\bs{x}}$ in the right hand are for the single-vector channel (\ref{Equ:D4}).

Let's $\varrho\rightarrow 0$ such that $N'= 1$
\begin{align}
\nonumber
&\mathbb{E}\left\{\Re(\bs{x}_n)^{\times i_{\text{R}}}\otimes \Im(\bs{x}_n)^{\times i_{\text{I}}}\otimes \Re(\hat{\bs{x}}_n)^{\times j_{\text{R}}}\otimes \Im(\hat{\bs{x}}_n)^{\times j_{\text{I}}}
\right\}\\
&=\mathbb{E}\left\{\Re(\bs{x})^{\times i_{\text{R}}}\otimes \Im(\bs{x})^{\times i_{\text{I}}}\otimes \Re(\hat{\bs{x}})^{\times j_{\text{R}}}\otimes \Im(\hat{\bs{x}})^{\times j_{\text{I}}} \right\}.
\end{align}
This completes the proof of equivalent joint moments, that is each row $\bs{x}_n$ in system (\ref{Equ:system}) and its corresponding component of MMSE estimator $\hat{\bs{x}}_n$ is decoupled into a set of ``parallel'' single-vector channels with the form of (\ref{Equ:D4}).

\section{Realization of MMSE Estimator: A Message Passing Based Algorithm}
\label{Sec:MP}
In this section, we propose a computational efficient message passing based algorithm to recover the signal of interest with IID row prior and further give the performance analysis to analyze its achievable MSE performance. Using EP projection, we obtain a new algorithm, which is close to the GAMP \cite{rangan2011generalized} and the hybrid generalized AMP (HyGAMP) \cite{rangan2017hybrid}. Compared to GAMP, the proposed algorithm can handle the generalized linear model with IID row prior and row-wise mapping channel with GAMP as its special case under $M=1$. On the other hand, the difference between the proposed algorithm and HyGAMP is twofold. Firstly, the signal of interest in \cite{rangan2017hybrid} is column vector where $N$ blocks are involved, while it is matrix form with IID row prior in this paper. Secondly, \cite{rangan2017hybrid} only considers the real-valued region and it is hard to extend to complex-valued region due to the Taylor expansion in complex-valued region, while our algorithm and the derivation using EP projection avoids Taylor expansion such that it can both cover real-valued region and complex-valued region smoothly.

\subsection{Proposed Algorithm}
As observed from Algorithm \ref{alg:Proposed}. The proposed algorithm can be divided into five steps: (\ref{R1})-(\ref{R2}), (\ref{R3})-(\ref{R4}), (\ref{R5})-(\ref{R6}), (\ref{R7})-(\ref{R8}), and (\ref{R9})-(\ref{R10}).
\begin{itemize}
\item
In the first step (\ref{R1})-(\ref{R2}), the parameter $\bs{z}_{\ell}^{(t)}$ and $ \bs{Q}_{\ell}^{(z,t)}$ are calculated which construct the approximated prior $\mathcal{N}_c(\bs{z}_{\ell}|\bs{z}_{\ell}^{(t)}, \bs{Q}_{\ell}^{(z,t)})$ of $\bs{z}_{\ell}$. Obviously, this step only involves matrix multiple operation. Thus, the computational cost of this step is $\mathcal{O}(NL+M^2L)$.
\item
In the second step (\ref{R3})-(\ref{R4}), the MMSE estimator of $\bs{z}_{\ell}$ is performed. The posterior distribution $\mathcal{P}(\bs{z}_{\ell}|\bs{y}_{\ell})$ is approximated by the input approximated prior $\mathcal{N}_c(\bs{z}_{\ell}|\bs{z}_{\ell}^{(t)}, \bs{Q}_{\ell}^{(z,t)})$ and the known likelihood function $\mathcal{P}(\bs{y}_{\ell}|\bs{z}_{\ell})$ of $\bs{z}_{\ell}$. It is worth noting that this step does not change the dimension of input and its complexity cost is $\mathcal{O}(M^2L)$.
\item
In the third step (\ref{R5})-(\ref{R6}), the parameters $\bs{Q}^{(s,t)}_{\ell}$ and $\bs{s}_{\ell}^{(t)}$ are calculated. The main computational cost of this step is dominated by matrix inverse operation with the cost $\mathcal{O}(M^3)$. The complexity of this step is $\mathcal{O}(M^3L)$.
\item
In the fourth step (\ref{R7})-(\ref{R8}), the likelihood $\mathcal{P}(\bs{y}_{\ell}|\bs{X})$ is approximated by $\mathcal{N}_c(\bs{x}_{n}|\bs{r}_{n}^{(t)}, \bs{Q}_{n}^{(r,t)})$. The computational cost of this step is dominated by matrix inverse and matrix product, which is $\mathcal{O}(LN+M^2N)$.
\item
In the final step (\ref{R9})-(\ref{R10}), the MMSE estimator of $\bs{x}_{n}$ is carried out which is nonlinear except the simplest case such as Gaussian prior. Based on the input of approximated likelihood function $\mathcal{N}_c(\bs{x}_{n}|\bs{r}_{n}^{(t)}, \bs{Q}_{n}^{(r,t)})$ and known IID row prior, the posterior of each row of $\bs{X}$ is approximated by $\frac{\mathcal{P}_{\textsf{X}}(\bs{x}_n)\mathcal{N}_c(\bs{x}_{n}|\bs{r}_{n}^{(t)}, \bs{Q}_{n}^{(r,t)})}{\int \mathcal{P}_{\textsf{X}}(\bs{x}_n)\mathcal{N}_c(\bs{x}_{n}|\bs{r}_{n}^{(t)}, \bs{Q}_{n}^{(r,t)})\text{d}\bs{x}_n}$. As its byproduct, its mean and variance are evaluated, which further contributes to constructing the approximated prior of $\bs{z}_{\ell}$. Obviously, this step also does not change the dimension of input. Thus, the complexity of this step is $\mathcal{O}(M^2N)$. Meanwhile, $\hat{\bs{x}}_n^{(T)}$ and $\boldsymbol{\zeta}_{n}^{(T)}$ are the outputs of the proposed algorithm. More precisely, $\hat{\bs{x}}_n^{(T)}$ is point estimation while $\boldsymbol{\zeta}_{n}^{(T)}$ is statistical distribution.
\end{itemize}
To sum up, the complexity of the proposed algorithm is $\mathcal{O}(N^2+M^2N)$ by considering $N\uparrow \infty$ and fixed ratio $\alpha=\frac{L}{N}$.

\begin{figure}[!t]
\centering
\includegraphics[width=0.25\textwidth]{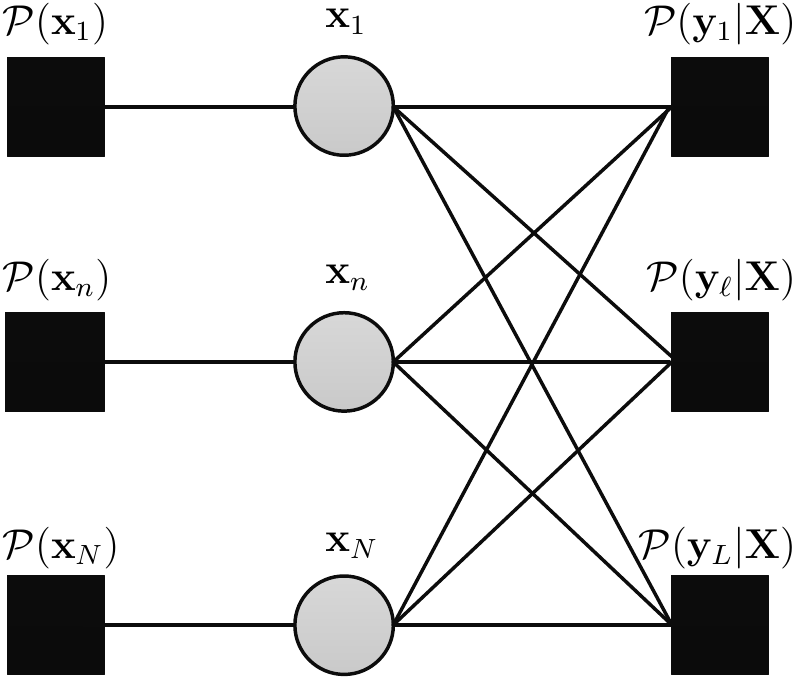}
\caption{The factor graph of $\mathcal{P}(\bs{X}, \bs{Y})$. The gray circles denote variable nodes, while black squares refer to factor nodes. The messages update via the edges between variable nodes and factor nodes.
}
\label{fig:FG}
\end{figure}

\begin{algorithm}[!t]
\caption{Proposed algorithm}
\label{alg:Proposed}
{
\begingroup
\textbf{1. Definition:}
\begin{align}
\boldsymbol{\zeta}_{\ell}^{(t)}&\sim \frac{\mathcal{P}(\bs{y}_{\ell}|\bs{z}_{\ell})\mathcal{N}_c(\bs{z}_{\ell}|\bs{z}_{\ell}^{(t)}, \bs{Q}_{\ell}^{(z,t)})}{\int \mathcal{P}(\bs{y}_{\ell}|\bs{z}_{\ell})\mathcal{N}_c(\bs{z}_{\ell}|\bs{z}_{\ell}^{(t)}, \bs{Q}_{\ell}^{(z,t)})\text{d}\bs{z}_{\ell}}
\tag{D1} \label{D1}\\
\boldsymbol{\xi}_n^{(t+1)}&\sim \frac{\mathcal{P}_{\textsf{X}}(\bs{x}_n)\mathcal{N}_c(\bs{x}_n|\bs{r}_n^{(t)}, \bs{Q}_n^{(r,t)})}{\int \mathcal{P}_{\textsf{X}}(\bs{x}_n)\mathcal{N}_c(\bs{x}_n|\bs{r}_n^{(t)}, \bs{Q}_n^{(r,t)})\text{d}\bs{x}_n}
\tag{D2} \label{D2}
\end{align}
\textbf{2. Initialization:} Choosing $\hat{\bs{Q}}_n^{(x,1)}$, $\hat{\bs{x}}_n^{(1)}$, $\bs{s}_{\ell}^{(0)}$.\\
\textbf{3. Output:} $\hat{\bs{X}}$, $\{\boldsymbol{\zeta}_{n}^{(t)}, \forall n \}$. \\
\textbf{4. Iteration:}\\
\For{$t=1,\cdots,T$}
{
    \setlength\abovedisplayskip{0pt}
    \setlength\belowdisplayskip{0pt}
    \For{$\ell=1,\cdots,L$}
    {
        \setlength\abovedisplayskip{0pt}
        \setlength\belowdisplayskip{0pt}
        \begin{align}
        \bs{Q}_{\ell}^{(z,t)}&=\sum_{n=1}^N |h_{\ell n}|^2\hat{\bs{Q}}_n^{(x,t)}
        \tag{R1} \label{R1}\\
        \bs{z}_{\ell}^{(t)}&=\sum_{n=1}^N h_{\ell n}\hat{\bs{x}}_n^{(t)}-\bs{Q}_{\ell}^{(z,t)}\bs{s}_{\ell}^{(t-1)}
        \tag{R2} \label{R2}\\
        \tilde{\bs{z}}_{\ell}^{(t)}&=\mathbb{E}_{\boldsymbol{\zeta}_{\ell}^{(t)}}\left\{\boldsymbol{\zeta}_{\ell}^{(t)}\right\}
        \tag{R3} \label{R3}\\
        \tilde{\bs{Q}}_{\ell}^{(z,t)}&=\text{Var}_{\boldsymbol{\zeta}_{\ell}^{(t)}}\left\{\boldsymbol{\zeta}_{\ell}^{(t)}\right\}
        \tag{R4} \label{R4}\\
        \bs{Q}_{\ell}^{(s,t)}&=(\bs{Q}_{\ell}^{(z,t)})^{-1}(\bs{Q}_{\ell}^{(z,t)}-\tilde{\bs{Q}}_{\ell}^{(z,t)})(\bs{Q}_{\ell}^{(z,t)})^{-1}
        \tag{R5} \label{R5}\\
        \bs{s}_{\ell}^{(t)}&=(\bs{Q}_{\ell}^{(z,t)})^{-1}(\tilde{\bs{z}}_{\ell}^{(t)}-\bs{z}_{\ell}^{(t)})
        \tag{R6} \label{R6}
        \end{align}
    }
    \For{$n=1,\cdots,N$}
    {
        \setlength\abovedisplayskip{0pt}
        \setlength\belowdisplayskip{0pt}
        \begin{align}
        \bs{Q}_n^{(r,t)}&=\left(\sum_{\ell=1}^L |h_{\ell n}|^2\bs{Q}_{\ell}^{(s,t)}\right)^{-1}
        \tag{R7} \label{R7}\\
        \bs{r}_{n}^{(t)}&=\hat{\bs{x}}_n^{(t)}+\bs{Q}_n^{(r,t)}\sum_{\ell=1}^Lh_{\ell n}^{*}\bs{s}_{\ell}^{(t)}
        \tag{R8} \label{R8}\\
        \hat{\bs{x}}_n^{(t+1)}&=\mathbb{E}_{\boldsymbol{\xi}_n^{(t+1)}}\left\{\boldsymbol{\xi}_n^{(t+1)}\right\}
        \tag{R9} \label{R9}\\
        \hat{\bs{Q}}_n^{(x,t+1)}&=\text{Var}_{\boldsymbol{\xi}_n^{(t+1)}}\left\{\boldsymbol{\xi}_n^{(t+1)}\right\}
        \tag{R10} \label{R10}
        \end{align}
    }
}
\endgroup
}
\end{algorithm}

\subsection{Derivation using EP}
We begin at introducing the factor graph shown in Fig.~\ref{fig:FG} corresponding to the joint distribution $\mathcal{P}(\bs{X},\bs{Y})$\footnote{
In practical algorithm, we estimate the $\bs{X}$ from $\bs{Y}$ given $\bs{H}$. So the conditional distribution $\mathcal{P}(\bs{Y}|\bs{H},\bs{X})$ given $\bs{H}$ is abbreviated as $\mathcal{P}(\bs{Y}|\bs{X})$, while it is $\mathcal{P}(\bs{Y}|\bs{H},\bs{X})$ in replica analysis since we analyze the asymptotic performance over observation $\bs{Y}$ and measurement matrix $\bs{H}$.
} with the form of
\begin{align}
\nonumber
\mathcal{P}(\bs{X},\bs{Y})&=\mathcal{P}(\bs{X})\mathcal{P}(\bs{Y}|\bs{X})\\
&=\prod_{n=1}^N \mathcal{P}_{\textsf{X}}(\bs{x}_n)\prod_{\ell=1}^L \mathcal{P}(\bs{y}_{\ell}|\bs{X}).
\end{align}

Using EP projection rules \cite[Fig.~2]{zou2018concise}\cite{meng2015concise}, the messages in Fig.~\ref{fig:FG} are addressed as
\begin{subequations}
\begin{align}
\mu_{n\leftarrow \ell}^{(t)}(\bs{x}_n)&\propto \frac{\text{Proj}\left[\mu_{n\rightarrow \ell}^{(t)}(\bs{x}_n)f_{n\leftarrow \ell}^{(t)}(\bs{x}_n)\right]}{\mu_{n\rightarrow \ell}^{(t)}(\bs{x}_n)}
\label{Equ:EP1},\\
\mu_{n\rightarrow \ell}^{t+1}(\bs{x}_n)&\propto \frac{\text{Proj}\left[\mathcal{P}_{\textsf{X}}(\bs{x}_n)\prod_{l}\mu_{n\leftarrow l}^{(t)}(\bs{x}_n)\right]}{\mu_{n\leftarrow \ell}^{(t)}(\bs{x}_n)}
\label{Equ:EP2},
\end{align}
\end{subequations}
where $\mu_{n\leftarrow \ell}^{(t)}(\bs{x}_n)$ is the message from factor node $\mathcal{P}(\bs{y}_{\ell}|\bs{X})$ to vector node $\bs{x}_n$ while $\mu_{n\rightarrow \ell}^{(t)}(\bs{x}_n)$ is the message in opposite direction. In addition, the term $f_{n\leftarrow \ell}^{(t)}(\bs{x}_n)$ is defined as
\begin{align}
f_{n\leftarrow \ell}^{(t)}(\bs{x}_n)=\int \mathcal{P}(\bs{y}_{\ell}|\bs{X})\prod_{i\ne n}\mu_{i\rightarrow \ell}^{(t)}(\bs{x}_i)\text{d}\bs{X}_{\backslash n},
\end{align}
where $\bs{X}_{\backslash n}$ is $\bs{X}$ except $\bs{x}_n$. The essence of EP projection is to map a distribution to Gaussian family and let `$\text{Proj}$' represent this operation. It says $\text{Proj}[p(\bs{x})]=\mathcal{N}_c(\bs{x}|\bs{a}, \bs{A})$ with $\bs{a}=\int \bs{x}p(\bs{x})\text{d}\bs{x}$ and $\bs{A}=\int (\bs{x}-\bs{a})(\bs{x}-\bs{a})^{\dag}p(\bs{x})\text{d}\bs{x}$. The detailed procedures are shown in Appendix \ref{App:EP}.

\subsection{Asymptotic Performance}
Inspired by self-averaged property, in which the affine of random variables $|f(\bs{y},\bs{A})-\mathbb{E}_{\bs{y},\bs{A}}\{f(\bs{y},\bs{A})\}|$ tends to vanish in large system limit, it assume
\begin{align}
\hat{\bs{Q}}_n^{(x,t)}&\approx \frac{1}{N}\sum_{n=1}^N\hat{\bs{Q}}_n^{(x,t)}=\overline{\hat{\bs{Q}}^{(x,t)}},\\
\tilde{\bs{Q}}_{\ell}^{(z,t)}&\approx \frac{1}{L}\sum_{\ell=1}^L\tilde{\bs{Q}}_{\ell}^{(z,t)}=\overline{\tilde{\bs{Q}}^{(z,t)}}.
\end{align}
Based on this assumption
\begin{align}
\bs{Q}_{\ell}^{(z,t)}&\approx \frac{1}{\alpha}\overline{\hat{\bs{Q}}^{(x,t)}}=\bs{Q}^{(z,t)},\\
\nonumber
\bs{Q}_{\ell}^{(s,t)}&\approx (\bs{Q}^{(z,t)})^{-1}(\bs{Q}^{(z,t)}-\tilde{\bs{Q}}^{(z,t)})(\bs{Q}^{(z,t)})^{-1}\\
                     &=\bs{Q}^{(s,t)},\\
\bs{Q}_n^{(r,t)}&\approx (\bs{Q}^{(s,t)})^{-1}.
\end{align}
Before giving the detailed procedures of performance analysis, let's introduce the following concepts:
\begin{definition}[Pseudo-lipschitz function]
For any $k\geq 1$, a function $\varphi:\mathbb{C}^p\mapsto \mathbb{C}^q$ is pseudo-Lipschitz of order $k$, if there exists a constant $C>0$ such that for any $\bs{x}, \bs{y}\in \mathbb{C}^p$,
\begin{align}
|\varphi(\bs{x})-\varphi(\bs{y})|\leq C(1+\|\bs{x}\|^{k-1}+\|\bs{y}\|^{k-1})\|\bs{x}-\bs{y}\|.
\end{align}
\end{definition}

\begin{definition}[Empirically converges to random vector]
Let $\bs{X}=\{\bs{x}_n(N)\}_{n=1}^N$ be a block sequence set with $\bs{x}_n(N)\in \mathbb{C}^p \ (p\geq 1)$. Given $k\geq 1$, $\bs{X}$ converges empirically a random vector $\textsf{X}$ on $\mathbb{C}^p$ with $k$-order moments if:\\
(i) $\mathbb{E}\{\|\textsf{X}\|^k\}<\infty$; and \\
(ii) for any pseudo-Lipschitz continuous function $\varphi(\cdot)$ of order $k$; then
\begin{align}
\lim_{N\rightarrow \infty}\frac{1}{N}\sum_{n=1}^N\varphi(\bs{x}_n(N))-\mathbb{E}\{\varphi(\textsf{X})\}\overset{\rm{a.s.}}{=}\bs{0},
\end{align}
We say $\varphi (\bs{x}_n)$ converges to $\mathbb{E}\{\varphi(\textsf{X})\}$ and write it as $\lim_{N\rightarrow \infty}\{\bs{x}_n(N)\}\overset{PL(k)}{=}\textsf{X}$, where $PL$ is the abbreviation of  Pseudo-lipschitz.
\end{definition}

We now make the following assumption:
\begin{assumption}
\label{Assum:1}
It is assumed that the mean-related parameters $\bs{r}_n^{(t)}$, $\bs{z}_{\ell}$, $\bs{z}_{\ell}^{(t)}$, $\bs{y}_{\ell}$ converge empirically to random vectors with 2nd order moments:
\begin{align}
\lim_{L,N\rightarrow \infty}\left\{\bs{r}_n^{(t)}, \bs{z}_{\ell}, \bs{z}_{\ell}^{(t)}, \bs{y}_{\ell}\right\}\overset{PL(2)}{=}\{\textsf{r}^{(t)},  \textsf{z}, \textsf{z}^{(t)}, \textsf{y}\}.
\end{align}
\end{assumption}

Note that we here pay close attention to the asymptotic MSE performance of proposed algorithm.  The asymptotic MSE of proposed algorithm of $t$-iteration is defined by
\begin{align}
\nonumber
\!\!\!\textsf{MSE}(\bs{X},t)
&=\lim_{N\rightarrow \infty}\frac{1}{NM}\|\bs{X}-\hat{\bs{X}}^{(t)}\|_{\text{F}}^2\\
\nonumber
&=\frac{1}{M}\frac{1}{N}\sum_{n=1}^N\text{Tr}\left((\hat{\bs{x}}_n^{(t)}-\bs{x}_n)(\hat{\bs{x}}_n^{(t)}-\bs{x}_n)^{\dag}\right)\\
&\overset{\text{a.s.}}{=}\frac{1}{M}\text{Tr}\left(\mathbb{E}\left\{(g(\textsf{X},\textsf{r}^{(t)})-\textsf{X})(g(\textsf{X},\textsf{r}^{(t)})-\textsf{X})^{\dag}\right\}\right)
\label{Equ:E5}\\
&\overset{\text{a.s.}}{=}\frac{1}{M}\text{Tr}\left(\overline{\hat{\bs{Q}}^{(x,t)}}\right),
\label{Equ:E6}
\end{align}
where $\bs{x}_n\sim \mathcal{P}_{\textsf{X}}(\bs{x}_n)$, and $g(\bs{x}, \bs{r}^{(t)})$ is associated with $\hat{\bs{x}}^{(t)}$ with the form of
\begin{align}
g(\bs{x}, \bs{r}^{(t)})=\frac{\int \bs{x}\mathcal{P}_{\textsf{X}}(\bs{x})\mathcal{N}_c(\bs{x}|\bs{r}^{(t)}, \bs{Q}^{(r,t)})\text{d}\bs{x}}{\int \mathcal{P}_{\textsf{X}}(\bs{x})\mathcal{N}_c(\bs{x}|\bs{r}^{(t)}, \bs{Q}^{(r,t)})\text{d}\bs{x}},
\end{align}
and the expectation in (\ref{Equ:E5}) is over $\mathcal{P}(\bs{x},\bs{r}^{(t)})=\mathcal{P}_{\textsf{X}}(\bs{x})\mathcal{N}_c(\bs{x}|\bs{r}^{(t)}, \bs{Q}^{(r,t)})$. It is worthy of noting that the distribution $\mathcal{N}_c(\bs{x}|\bs{r}^{(t)}, \bs{Q}^{(r,t)})$ is the  approximated likelihood function and $\bs{r}^{(t)}$ is mean parameter that relates to the observation. Thus, $\mathcal{N}_c(\bs{x}|\bs{r}^{(t)}, \bs{Q}^{(r,t)})$ is interpreted as $\mathcal{P}(\bs{r}^{(t)}|\bs{x})$. Sequentially, the averaged covariance matrix $\overline{\hat{\bs{Q}}^{(x,t)}}$ can be expressed as
\begin{align}
\nonumber
\overline{\hat{\bs{Q}}^{(x,t)}}
&=\mathbb{E}_{\textsf{X}}\left\{\textsf{X}\textsf{X}^{\dag}\right\}-\mathbb{E}_{\textsf{r}^{(t)}}\left\{g(\textsf{X},\textsf{r}^{(t)})g(\textsf{X},\textsf{r}^{(t)})^{\dag}\right\}\\
&=\bs{\Xi}_x-\int \frac{\bs{U}_{\bs{x}}(\boldsymbol{\zeta}, \bs{Q}^{(r,t)})}{v_{\bs{x}}(\boldsymbol{\zeta}, \bs{Q}^{(r,t)})}{\rm{d}}\boldsymbol{\zeta}.
\end{align}

From (\ref{Equ:E6}), it claims that the asymptotic MSE of proposed algorithm is identical to the averaged trace of $\overline{\hat{\bs{Q}}^{(x,t)}}$. In addition, it can be found that only the covariance-related parameters $\bs{Q}^{(z,t)}$, $\overline{\tilde{\bs{Q}}^{(z,t)}}$, $\bs{Q}^{(s,t)}$ and $\tilde{\bs{Q}}^{(z,t)}$  have a relation to $\overline{\hat{\bs{Q}}^{(x,t)}}$. Among them, the relation between $\overline{\tilde{\bs{Q}}^{(z,t)}}$ and $\overline{\hat{\bs{Q}}^{(x,t)}}$ should be determined.

Based on the Assumption \ref{Assum:1}, we have
\begin{align}
\overline{\tilde{\bs{Q}}^{(z,t)}}
=\frac{1}{L}\sum_{\ell=1}^L \tilde{\bs{Q}}_{\ell}^{(z,t)}
\overset{\text{a.s.}}{=}\mathbb{E}_{\textsf{y}, \textsf{z}^{(t)}}\left\{\varphi(\textsf{y}, \textsf{z}^{(t)})\right\},
\label{Equ:F3}
\end{align}
where $\varphi(\bs{y}, \bs{z}^{(t)})$ has the form of
\begin{align}
\varphi(\bs{y}, \bs{z}^{(t)})
&=\frac{\int (*)\mathcal{P}(\bs{y}|\bs{z})\mathcal{N}_c(\bs{z}|\bs{z}^{(t)}, \bs{Q}^{(z,t)})\text{d}\bs{z}}{\int \mathcal{P}(\bs{y}|\bs{z})\mathcal{N}_c(\bs{z}|\bs{z}^{(t)}, \bs{Q}^{(z,t)})\text{d}\bs{z}},\\
(*)&=(\tilde{\bs{z}}^{(t)}-\bs{z})(\tilde{\bs{z}}^{(t)}-\bs{z})^{\dag}\\
\tilde{\bs{z}}^{(t)}&=\frac{\int \bs{z}\mathcal{P}(\bs{y}|\bs{z})\mathcal{N}_c(\bs{z}|\bs{z}^{(t)}, \bs{Q}^{(z,t)})\text{d}\bs{z}}
{\int \mathcal{P}(\bs{y}|\bs{z})\mathcal{N}_c(\bs{z}|\bs{z}^{(t)}, \bs{Q}^{(z,t)})\text{d}\bs{z}}.
\end{align}
Note that the joint distribution $\mathcal{P}(\bs{y}, \bs{z}^{(t)})$ is from $\mathcal{P}(\bs{y},\bs{z}^{(t)}, \bs{z})$. Also, we notice that the distribution $\mathcal{N}_c(\bs{z}|\bs{z}^{(t)}, \bs{Q}^{(z,t)})$ is the approximated prior of $\bs{z}$ given $\bs{z}^{(t)}$. As a result, this distribution can be interpreted as a transition distribution $\mathcal{P}(\bs{z}|\bs{z}^{(t)})$. Further, we have
\begin{align}
\nonumber
\mathcal{P}(\bs{y},\bs{z}^{(t)})&=\int \mathcal{P}(\bs{y},\bs{z}, \bs{z}^{(t)})\text{d}\bs{z}\\
&=\mathcal{P}(\bs{z}^{(t)})\int \mathcal{P}(\bs{y}|\bs{z})\mathcal{N}_c(\bs{z}|\bs{z}^{(t)}, \bs{Q}^{(z,t)})\text{d}\bs{z},
\label{Equ:F1}
\end{align}
where $\mathcal{P}(\bs{z}^{(t)})$ can be obtained by solving
\begin{align}
\int \mathcal{N}_c(\bs{z}|\bs{z}^{(t)}, \bs{Q}^{(z,t)})\mathcal{P}(\bs{z}^{(t)})\text{d}\bs{z}^{(t)}=\mathcal{P}(\bs{z})
\label{Equ:E7}.
\end{align}
In fact, since $\mathcal{P}(\bs{z})$ is unknown, solving this equation is very difficult. In large system limit, the central limit theorem allows us to treat $\bs{z}$ as a Gaussian vector with zero mean and covariance matrix
\begin{align}
\bs{Q}_z
=\mathbb{E}_{\bs{H},\bs{X}}\left\{\bs{z}\bs{z}^{\dag}\right\}
=\sum_{n=1}^N\mathbb{E}\left\{|h_{\ell n}|^2\bs{xx}^{\dag}\right\}
=\frac{1}{\alpha}\bs{\Xi}_x.
\end{align}
By Gaussian approximation $\mathcal{P}(\bs{z})=\mathcal{N}_c(\bs{z}|\bs{0},\frac{1}{\alpha}\bs{\Sigma}_x)$, it is easy to find the solution of (\ref{Equ:E7})
\begin{align}
\mathcal{P}(\bs{z}^{(t)})=\mathcal{N}_c(\bs{z}^{(t)}|\bs{0},\frac{1}{\alpha}\bs{\Xi}_x-\bs{Q}^{(z,t)}).
\label{Equ:F2}
\end{align}
Taking (\ref{Equ:F1}) and (\ref{Equ:F2}) into $\overline{\tilde{\bs{Q}}^{(z,t)}}$ in (\ref{Equ:F3}), we obtain
\begin{align}
\nonumber
\overline{\tilde{\bs{Q}}^{(z,t)}}
&=\mathbb{E}_{\textsf{y}, \textsf{z}^{(t)}}\left\{
\mathbb{E}_{\textsf{z}|\textsf{z}^{(t)},\textsf{y}}\left\{\textsf{zz}^{\dag}\right\}-\mathbb{E}_{\textsf{z}|\textsf{z}^{(t)},\textsf{y}}\{\textsf{z}\}\mathbb{E}_{\textsf{z}|\textsf{z}^{(t)},\textsf{y}}\{\textsf{z}\}^{\dag}
\right\}\\
\nonumber
&=\frac{1}{\alpha}\bs{\Xi}_x-\int \frac{\int \bs{z}\mathcal{P}(\bs{y}|\bs{z})\mathcal{N}_c(\bs{z}|\bs{z}^{(t)}, \bs{Q}^{(z,t)})\text{d}\bs{z}}{
\int \mathcal{P}(\bs{y}|\bs{z})\mathcal{N}_c(\bs{z}|\bs{z}^{(t)}, \bs{Q}^{(z,t)})\text{d}\bs{z}
}\\
\nonumber
&\qquad \qquad\times \left(\int \bs{z}\mathcal{P}(\bs{y}|\bs{z})\mathcal{N}_c(\bs{z}|\bs{z}^{(t)}, \bs{Q}^{(z,t)})\text{d}\bs{z}\right)^{\dag}\\
&\qquad \qquad \times \mathcal{N}_c(\bs{z}^{(t)}|\bs{0},\frac{1}{\alpha}\bs{\Sigma}_x-\bs{Q}^{(z,t)})\text{d}\bs{z}^{(t)}\text{d}\bs{y}.
\label{Equ:F4}
\end{align}
Using the fact $\bs{Q}^{(z,t)}=\frac{1}{\alpha}\overline{\hat{\bs{Q}}^{(x,t)}}$ and defining $\bs{z}^{(t)}=\frac{1}{\sqrt{\alpha}}(\bs{\Sigma}_x-\overline{\hat{\bs{Q}}^{(x,t)}})^{\frac{1}{2}}\boldsymbol{\zeta}$, (\ref{Equ:F4}) becomes
\begin{align}
\nonumber
&\overline{\tilde{\bs{Q}}^{(z,t)}}
=\frac{1}{\alpha}\bs{\Xi}_x\\
&\quad -\int \frac{\bs{U}_{\bs{y}|\bs{z}}\left(\frac{1}{\sqrt{\alpha}}(\bs{\Sigma}_x-\overline{\hat{\bs{Q}}^{(x,t)}})^{\frac{1}{2}}\boldsymbol{\zeta}, \frac{1}{\alpha}\overline{\hat{\bs{Q}}^{(x,t)}}\right)}
{v_{\bs{y}|\bs{z}}\left(\frac{1}{\sqrt{\alpha}}(\bs{\Sigma}_x-\overline{\hat{\bs{Q}}^{(x,t)}})^{\frac{1}{2}}\boldsymbol{\zeta}, \frac{1}{\alpha}\overline{\hat{\bs{Q}}^{(x,t)}}\right)}\text{\rm{D}}\boldsymbol{\zeta}{\rm{d}}\bs{y}.
\end{align}

Summarily, we build a close loop where only the asymptotic MSE-related parameters are involved, and those equations are called SE which predicts the dynamics of the proposed algorithm. For convenience, we post them as below
\begin{subequations}
\begin{align}
\nonumber
\!\!\!\!\!\!&\overline{\tilde{\bs{Q}}^{(z,t)}}
=\frac{1}{\alpha}\bs{\Xi}_x\\
\!\!\!\!\!\!&-
\int \frac{\bs{U}_{\bs{y}|\bs{z}}\left(\frac{1}{\sqrt{\alpha}}(\bs{\Sigma}_x-\overline{\hat{\bs{Q}}^{(x,t)}})^{\frac{1}{2}}\boldsymbol{\zeta}, \frac{1}{\alpha}\overline{\hat{\bs{Q}}^{(x,t)}}\right)}
{v_{\bs{y}|\bs{z}}\left(\frac{1}{\sqrt{\alpha}}(\bs{\Sigma}_x-\overline{\hat{\bs{Q}}^{(x,t)}})^{\frac{1}{2}}\boldsymbol{\zeta}, \frac{1}{\alpha}\overline{\hat{\bs{Q}}^{(x,t)}}\right)}\text{\rm{D}}\boldsymbol{\zeta}{\rm{d}}\bs{y},\\
\!\!\!\!\!\!&\bs{Q}^{(r,t)}=\overline{\hat{\bs{Q}}^{(x,t)}}\left(\alpha\overline{\hat{\bs{Q}}^{(x,t)}}-\alpha^2\overline{\tilde{\bs{Q}}^{(z,t)}}\right)^{-1}
\overline{\hat{\bs{Q}}^{(x,t)}},\\
\!\!\!\!\!\!&\overline{\hat{\bs{Q}}^{(x,t)}}=\bs{\Xi}_x-\int \frac{\bs{U}_{\bs{x}}(\boldsymbol{\zeta}, \bs{Q}^{(r,t)})}{v_{\bs{x}}(\boldsymbol{\zeta}, \bs{Q}^{(r,t)})}{\rm{d}}\boldsymbol{\zeta}.
\end{align}
\label{Equ:SE}
\end{subequations}
Comparing (\ref{Equ:SE}) with (\ref{SD}), under $\bs{Q}^{(r,t)}=\tilde{\bs{B}}^{-1}$, the SE of proposed algorithm shares the same equations with the fixed point (\ref{SD}) of the exact MMSE estimator as predicted by replica method. That illustrates, the proposed algorithm can attain the Bayes-optimal MSE error if its fixed point is unique.

\section{Simulation and Discussion}
In this section, we give the numeric simulations to verify the accuracy of our analytical results and show the performance of the proposed algorithm. The signal-to-noise (SNR) and  normalized MSE (NMSE) of $\bs{X}$ are  define as $\frac{\mathbb{E}\{\|\bs{HX}\|_{\text{F}}^2\}}{\mathbb{E}\{\|\bs{W}\|_{\text{F}}^2\}}$
and $\frac{\mathbb{E}\{\|\bs{X}-\hat{\bs{X}}\|_{\text{F}}^2\}}{\mathbb{E}\{\|\bs{X}\|_{\text{F}}^2\}}$, respectively.

\subsection{Quantized Compressed Sensing}
In this subsection, we consider the application of the objective model (\ref{Equ:system}) in quantized compressed sensing, which can be modeled as
\begin{align}
\bs{Y}=\textsf{Q}_{\textsf{c}}(\bs{HX}+\bs{W}),
\label{Equ:Compressed}
\end{align}
where $\bs{X}\in \mathbb{C}^{N\times M}$ is row-sparse in which each row $\bs{x}_n\in \mathbb{C}^M$ is assumed to be drawn from $\mathcal{P}_{\textsf{X}}(\bs{x}_n)$ with the form of
\begin{align}
\mathcal{P}_{\textsf{X}}(\bs{x}_n)=\rho\mathcal{N}_c(\bs{x}_n|\bs{0}, \bs{\Sigma}_x)+(1-\rho)\delta(\bs{x}_n),
\label{Equ:prior}
\end{align}
In addition, it is assumed that each entry of $\bs{H}\in \mathbb{C}^{L\times N}$ is drawn from Gaussian distribution with zero mean and $1/L$ variance, and each row $\bs{w}_n$ of $\bs{W}$ follows $\mathcal{N}_c(\bs{w}_n|\bs{0}, \bs{\Sigma}_w)$. $\textsf{Q}_{\text{c}}$ denotes a complex-valued and uniformed quantizer involving two separable real-valued quantizer, and the setups of quantization follows \cite{wen2015bayes}. In this case, the transition distribution is given by
$
\mathcal{P}(\bs{y}_{\ell}|\bs{z}_{\ell})=\mathcal{P}(\Re(\bs{y}_{\ell})|\Re(\bs{z}_{\ell}))\mathcal{P}(\Im({\bs{y}}_{\ell}))|\Im(\bs{z}_{\ell}))
$, where
$
\mathcal{P}(\Re(\bs{y}_{\ell})|\Re(\bs{z}_{\ell}))=\int_{q^{\text{low}}(\bs{y}_{\ell})}^{q^{\text{up}}(\bs{y}_{\ell})}\mathcal{N}(\bs{t}|\Re(\bs{z}_{\ell}), \frac{\bs{\Sigma}_w}{2})\text{d}\bs{t}
$, where $\int_{q^{\text{low}}(\bs{y}_{\ell})}^{q^{\text{up}}(\bs{y}_{\ell})}=\int_{q^{\text{low}}(\Re(y_{\ell 1}))}^{q^{\text{up}}(\Re(y_{\ell 1}))}\cdots \int_{q^{\text{low}}(\Re(y_{\ell M})}^{q^{\text{up}}(\Re(y_{\ell M})}$, $q^{\text{up}}(y)$ and $q^{\text{low}}(y)$ given by \cite[(64)-(65)]{zou2021multi} are the input range of quantization associated with $y$. $\mathcal{P}(\Im({\bs{y}}_{\ell}))|\Im(\bs{z}_{\ell}))$ is the same as $\mathcal{P}(\Re(\bs{y}_{\ell})|\Re(\bs{z}_{\ell}))$ by replacing $\Re$ with $\Im$. Specially, as the quantizer's bits tends to infinity, the system (\ref{Equ:Compressed}) becomes $\bs{Y}=\bs{HX}+\bs{W}$.

Given prior (\ref{Equ:prior}), by Gaussian reproduction property, the parameters $\hat{\bs{x}}_n^{(t+1)}, \hat{\bs{Q}}_n^{(x,t+1)}$ in Algorithm \ref{alg:Proposed} are given by
\begin{align}
\!\!\!\!\!\!\!\hat{\bs{x}}_n^{(t+1)}&=C_n^{(t)}\bs{p}_n^{(t)},\\
\!\!\!\!\!\!\!\hat{\bs{Q}}_n^{(x,t+1)}&=C_n^{(t)}(\bs{\Gamma}_n^{(t)}+\bs{p}_n^{(t)}(\bs{p}_n^{(t)})^{\dag})
-\hat{\bs{x}}_n^{(t+1)}(\hat{\bs{x}}_n^{(t+1)})^{\dag},
\end{align}
where $C_n^{(t)}=\frac{\rho \mathcal{N}_c(\bs{0}|\bs{r}_n^{(t)}, \bs{\Sigma}_x+\bs{Q}_{n}^{(r,t)})}{\rho \mathcal{N}_c(\bs{0}|\bs{r}_n^{(t)}, \bs{\Sigma}_x+\bs{Q}_{n}^{(r,t)})+(1-\rho)\mathcal{N}_c(\bs{0}|\bs{r}_n^{(t)},\bs{Q}_n^{(r,t)})}$, $\bs{\Gamma}_n^{(t)}=((\bs{Q}_n^{(r,t)})^{-1}+\bs{\Sigma}_x^{-1})^{-1}$, and $\bs{p}_n^{(t)}=\bs{\Gamma}_n^{(t)}(\bs{Q}_n^{(r,t)})^{-1}\bs{r}_n^{(t)}$. The $t$-iteration MMSE matrix $\textsf{MMSE}(\bs{X},t)$ of the proposed algorithm is given by
\begin{align}
\nonumber
&\textsf{MMSE}(\bs{X}, t)=\rho \bs{\Sigma}_x\\
&-\bs{\Lambda}^{(t)}\int \bs{rr}^{\dag}\frac{\rho \mathcal{N}_c(\bs{0}|\bs{r}, \bs{\Sigma}_x+\bs{Q}^{(r,t)})}{1+\frac{1-\rho}{\rho }\frac{\mathcal{N}_c(\bs{0}|\bs{r}, \bs{Q}^{(r,t)})}{\mathcal{N}_c(\bs{0}|\bs{r}, \bs{\Sigma}_x+\bs{Q}^{(r,t)})}}\text{d}\bs{r}\cdot (\bs{\Lambda}^{(t)})^{\dag},
\label{Equ:MMSE}
\end{align}
where $\bs{\Lambda}^{(t)}=\left[(\bs{Q}^{(r,t)})^{-1}+\bs{\Sigma}_x^{-1}\right]^{-1}(\bs{Q}^{(r,t)})^{-1}$. Note that, the evaluation of (\ref{Equ:MMSE}) can be obtained numerically by randomly generating $\bs{x}\sim \rho \mathcal{N}_c(\bs{x}|\bs{0},\bs{\Sigma}_x)+(1-\rho)\delta(\bs{x})$ and $\bs{r}=\bs{x}+\bs{n}$ with $\bs{n}\sim \mathcal{N}_c(\bs{0},\bs{Q}^{(r,t)})$.

In the case of infinite bits, by Gaussian reproduction property, the parameters $(\tilde{\bs{z}}_{\ell}, \tilde{\bs{Q}}_{\ell}^{(z,\ell)})$ are given by
\begin{align}
\tilde{\bs{Q}}_{\ell}^{(z,t)}&=\left(\bs{\Sigma}_w^{-1}+(\bs{Q}_{\ell}^{(z,t)})^{-1}\right)^{-1},
\label{Equ:Q}\\
\tilde{\bs{z}}_{\ell}^{(t)}&=\tilde{\bs{Q}}_{\ell}^{(z,t)}\left(\bs{\Sigma}_w^{-1}\bs{y}_{\ell}+(\bs{Q}_{\ell}^{(z,t)})^{-1}\bs{z}_{\ell}^{(t)}\right).
\label{Equ:z}
\end{align}
Furthermore, the SE (\ref{Equ:SE}) of the proposed algorithm reduces to
\begin{align}
\bs{Q}^{(r,t)}=\bs{\Sigma}_w+\frac{1}{\alpha}\textsf{MMSE}(\bs{X},t).
\end{align}

In the case of finite bits, there are not analytical expressions of $\tilde{\bs{z}}_{\ell}$ and $\tilde{\bs{Q}}_{\ell}^{(z,\ell)}$, but they can be evaluated by numerical integration. If considering $\bs{\Sigma}_x=\sigma_x^2\bs{I}$ and $\bs{\Sigma}_x\bs{I}$, then entries of $\tilde{\bs{z}}_{\ell}$ and $\tilde{\bs{Q}}_{\ell}^{(z,\ell)}$ can be obtained by \cite[(23)-(24)]{wen2015bayes}.

\subsection{Numeric Results}

\begin{figure}[!t]
\centering
\includegraphics[width=0.45\textwidth]{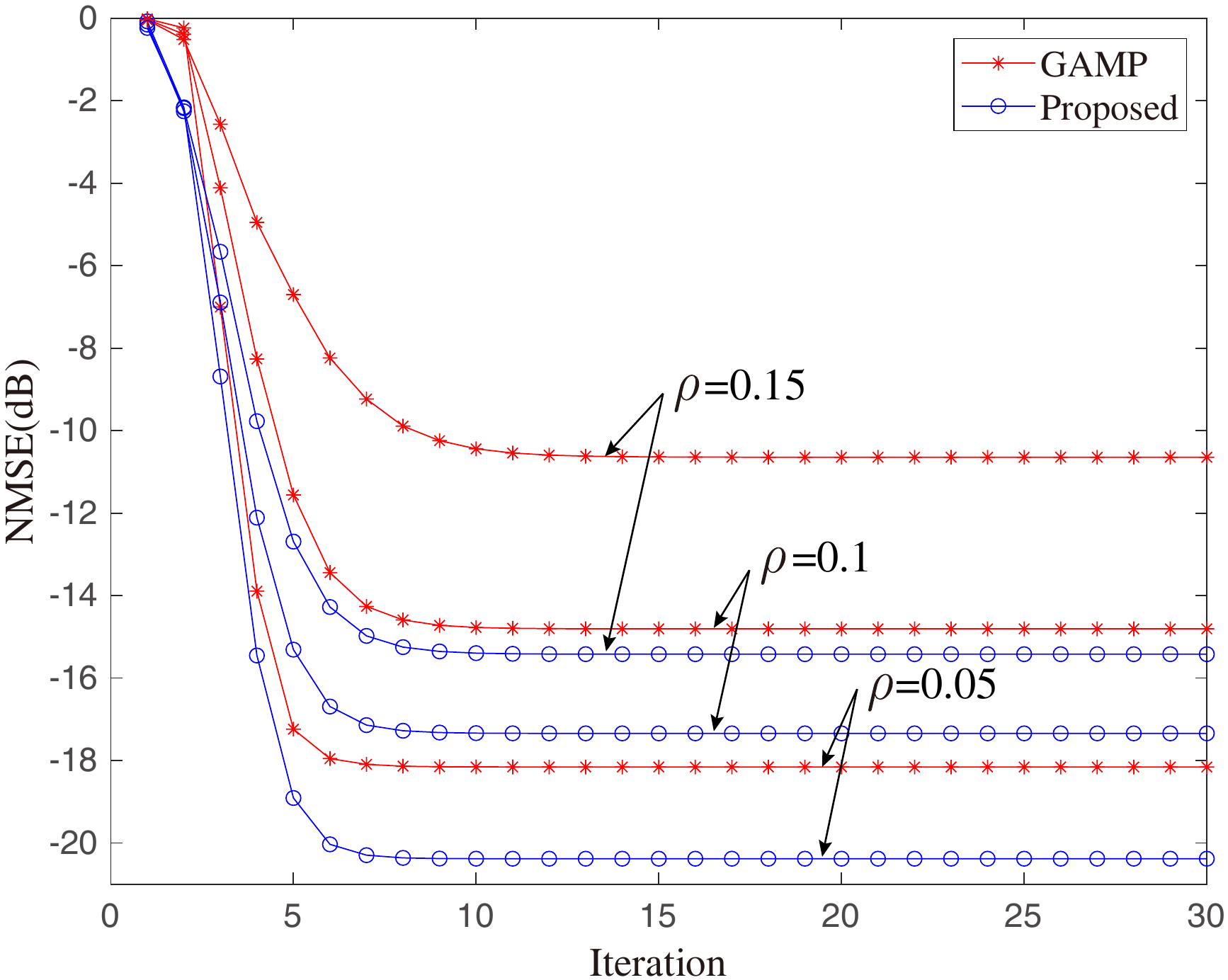}
\caption{Per-iteration NMSE performance of the proposed algorithm and GAMP. The dimensions are set as $(L,N,M)=(256, 512, 10)$ and SNR is 10dB. Each case of this experiment is over $10^4$ samplings.
}
\label{fig:Sim1}
\end{figure}

Fig.~\ref{fig:Sim1} compares the NMSE performance of the proposed algorithm with GAMP algorithm under the prior (\ref{Equ:prior}) and correlated Gaussian transition $\mathcal{N}_c(\bs{w}_n|\bs{0}, \bs{\Sigma}_w)$. The dimensions of system are set as $(L,N,M)=(256, 512, 10)$, and SNR is define as $\text{SNR}=10\log10\frac{\rho \text{Tr}(\bs{\Sigma}_x)}{\alpha\text{Tr}(\bs{\Sigma}_w)}$
and is set as 10dB. The covariance matrices $\bs{\Sigma}_x$ and $\bs{\Sigma}_w$ are generated by $\bs{\Sigma}_x=\frac{\bs{A}_x\bs{A}_x^{\dag}}{\text{Tr}(\bs{A}_x\bs{A}_x^{\dag})}$ and $\bs{\Sigma}_w=\frac{\rho 10^{-\frac{\text{SNR}}{10}}}{\alpha}\frac{\bs{A}_w\bs{A}_w^{\dag}}{\text{Tr}(\bs{A}_w\bs{A}_w^{\dag})}$ with $\bs{A}_x$ and $\bs{A}_w$ being random uniform matrices. Be aware, the GAMP algorithm runs using the diagonal elements of covariance-matrix, i.e., $\text{diag}(\bs{\Sigma}_x)$, $\text{diag}(\bs{\Sigma}_w)$. As can be seen from this figure, the proposed algorithm outperforms the GAMP algorithm in all settings (4.71dB gap in $\rho=0.15$). In addition, it can be found that the sparse ratio $\rho$ has the impact on both the speed of convergence and fixed point.  As sparse ratio $\rho$ increases, the speed of convergence of the proposed algorithm and GAMP decreases, and the NMSE performance  decreases.

\begin{figure}[!t]
\centering
\includegraphics[width=0.45\textwidth]{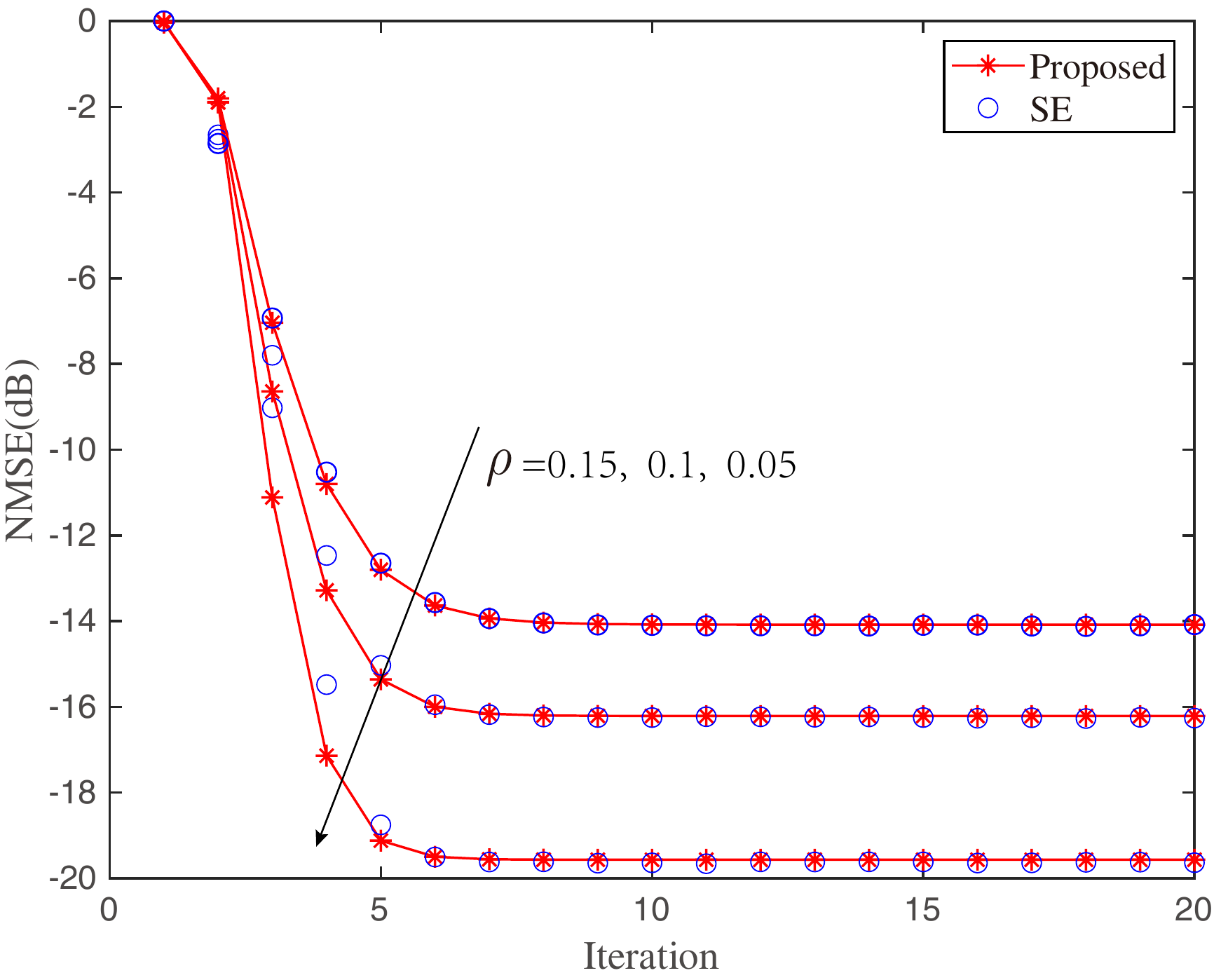}
\caption{Per-iteration NMSE of $\bs{X}$ comparing the proposed algorithm with its SE.
}
\label{fig:Sim3}
\end{figure}

Fig.~\ref{fig:Sim3} compares the NMSE performance of proposed algorithm and its SE\footnote{
Note that here we only present SE curves, since SE equations match perfectly with the replica formulas.}. The parameters are set as $(L, N, \rho)=(256, 512, 0.1)$. The covariance matrices $\bs{\Sigma}_x$ and $\bs{\Sigma}_w$ are generated by $\bs{\Sigma}_x=\frac{\bs{A}_x\bs{A}_x^{\dag}+2\bs{I}}{\text{Tr}(\bs{A}_x\bs{A}_x^{\dag}+2\bs{I})}$ and $\bs{\Sigma}_w=\frac{\rho 10^{-\frac{\text{SNR}}{10}}}{\alpha}\frac{\bs{A}_w\bs{A}_w^{\dag}+2\bs{I}}{\text{Tr}(\bs{A}_w\bs{A}_w^{\dag}+2\bs{I})}$ with $\bs{A}_x$ and $\bs{A}_w$ being random uniform matrices. It can be observed from this figure that SE tracks the proposed algorithm well in all settings.

\begin{figure}[!t]
\centering
\includegraphics[width=0.45\textwidth]{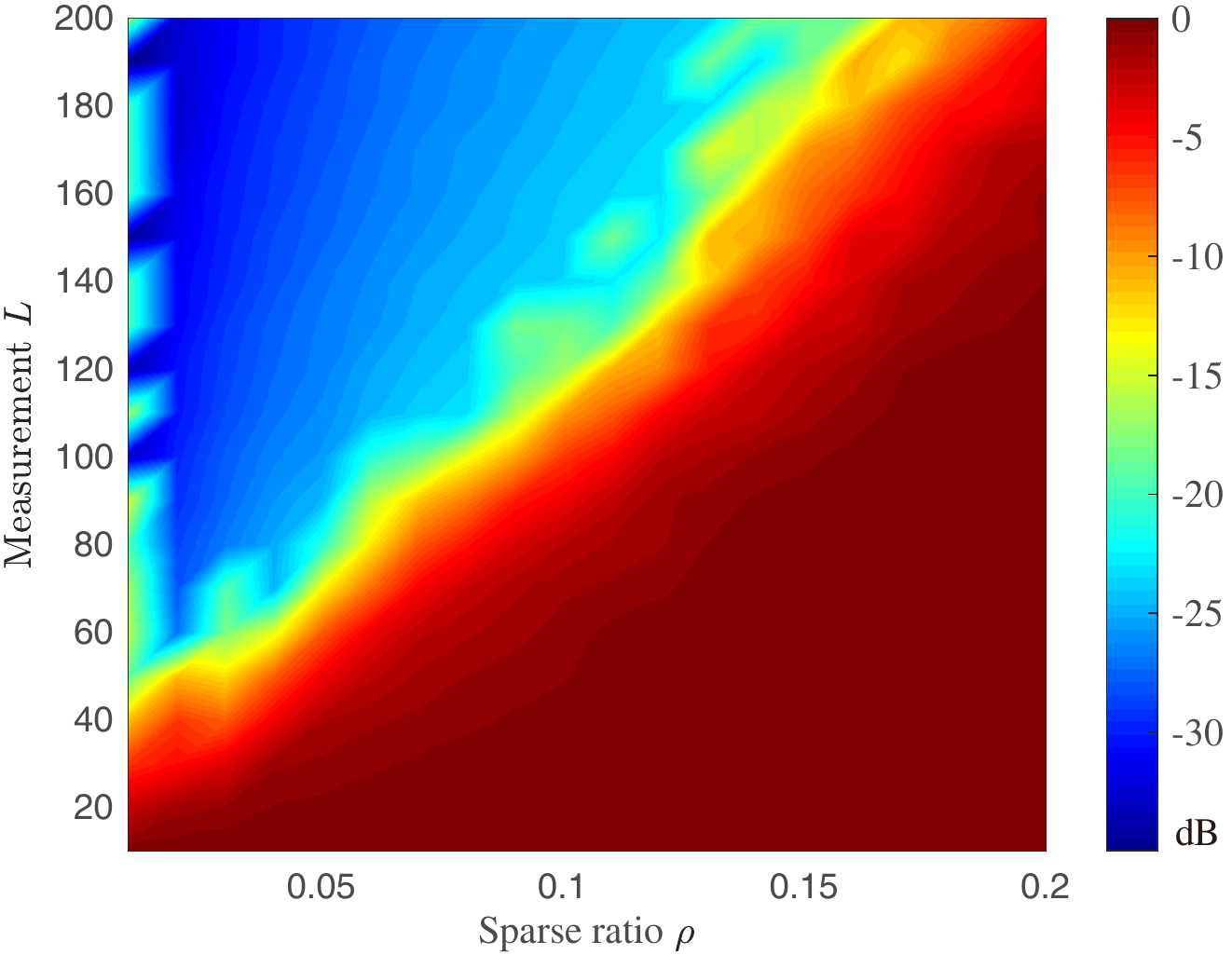}
\caption{Phase transition over a set of sparse ratios $\rho$ and measurements $L$, where we use color to denote NMSE of $\bs{X}$.
}
\label{fig:Sim2}
\end{figure}

Fig.~\ref{fig:Sim2} shows the phase transition of the proposed algorithm over a grid of sparse ratios and measurement ratios (by fixing $N=512$). The covariance matrices of $\bs{\Sigma}_x$ and $\bs{\Sigma}_w$ are generated by the same procedures as Fig.~\ref{fig:Sim1}. The SNR is set as 20dB and $M=20$. As can be observed from this figure, more measurement or smaller sparse ratio leads to a better NMSE.

\begin{figure}[!t]
\centering
\includegraphics[width=0.45\textwidth]{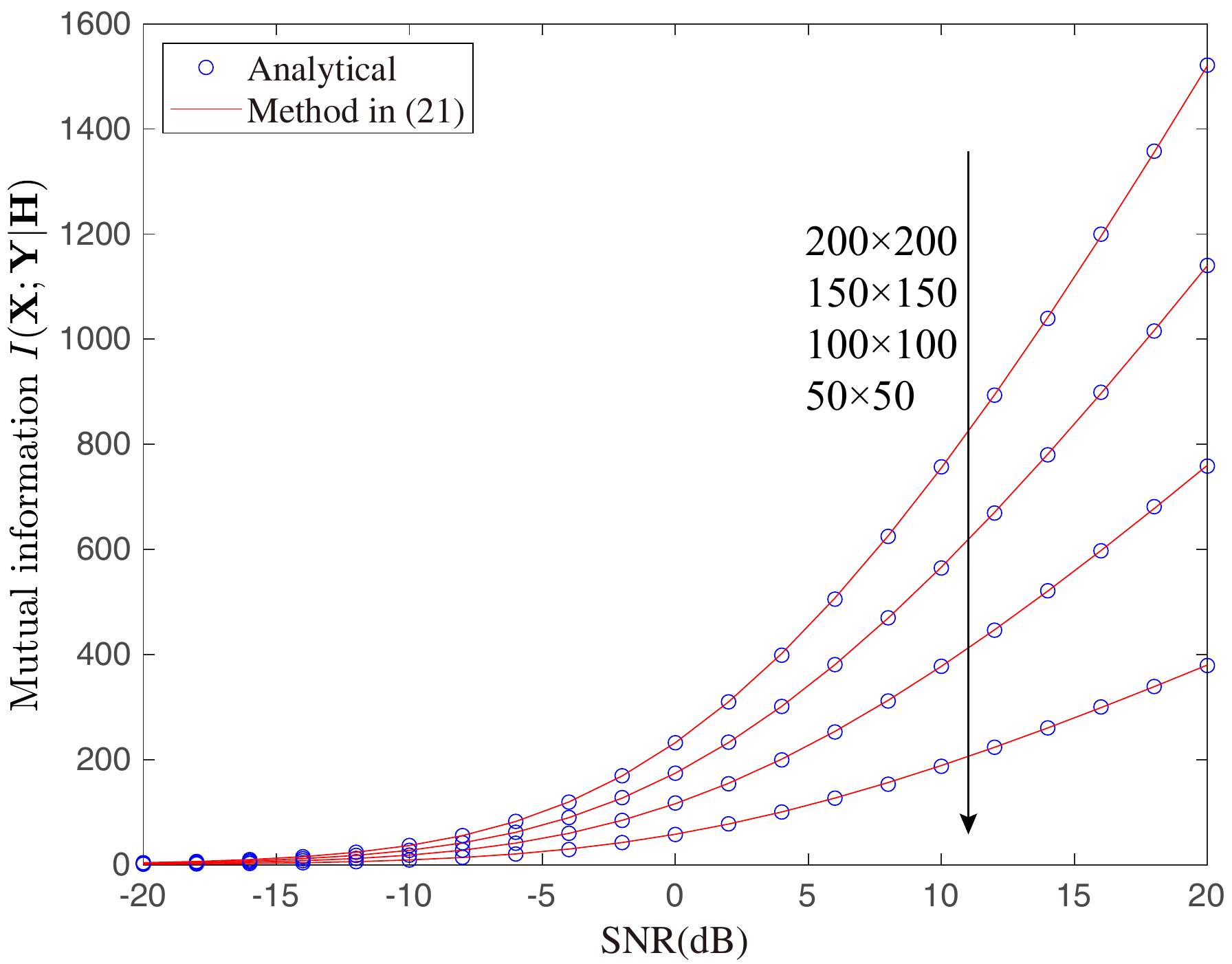}
\caption{Mutual information considering correlated Gaussian input, where $M=2$ is fixed and $L=N$.
}
\label{fig:Sim6}
\end{figure}

Considering Gaussian input $\mathcal{N}_c(\bs{x}_n|\bs{0}, \bs{\Sigma}_x)$, Fig.~\ref{fig:Sim6} depicts the mutual information by comparing  the method in (\ref{Mutual}) with the exact expression. The covariance matrices are set as $\bs{\Sigma}_x=\frac{\bs{I}+\bs{11}^{\text{T}}}{\text{Tr}(\bs{I}+\bs{11}^{\text{T}})}$, $\bs{\Sigma}_w=\frac{10^{-\frac{\text{SNR}}{10}}(\bs{I}+\bs{11}^{\text{T}})}{\alpha\text{Tr}(\bs{I}+\bs{11}^{\text{T}})}$. Given the Gaussian input, by Gaussian reproduction property, the mutual information is given by\footnote{
In large system limit, due to self-averaging property \cite{guo2005randomly}, the mutual information $I(\bs{X};\bs{Y})$ for an arbitrary realization  $\bs{H}$ converges to  $I(\bs{X};\bs{Y}|\bs{H})$.
 } $I(\bs{X};\bs{Y})=\log \det(\tilde{\bs{\Sigma}}_w+\tilde{\bs{H}}\tilde{\bs{\Sigma}}_x\tilde{\bs{H}}^{\dag})-\log\det(\tilde{\bs{\Sigma}}_w)$, where $\tilde{\bs{\Sigma}}_x=\bs{\Sigma}_x\otimes \bs{I}_N$, $\tilde{\bs{\Sigma}}_w=\bs{\Sigma}_w\otimes \bs{I}_L$, and $\tilde{\bs{H}}=\bs{H}\otimes \bs{I}_M$. As can be seen from this figure, our method matches perfectly with the analytical expression. However, in general, there may be not  the analytical expression of $I(\bs{X};\bs{Y}|\bs{H})$. In those cases, our method provides an alternative.

\begin{figure}[!t]
\centering
\includegraphics[width=0.43\textwidth]{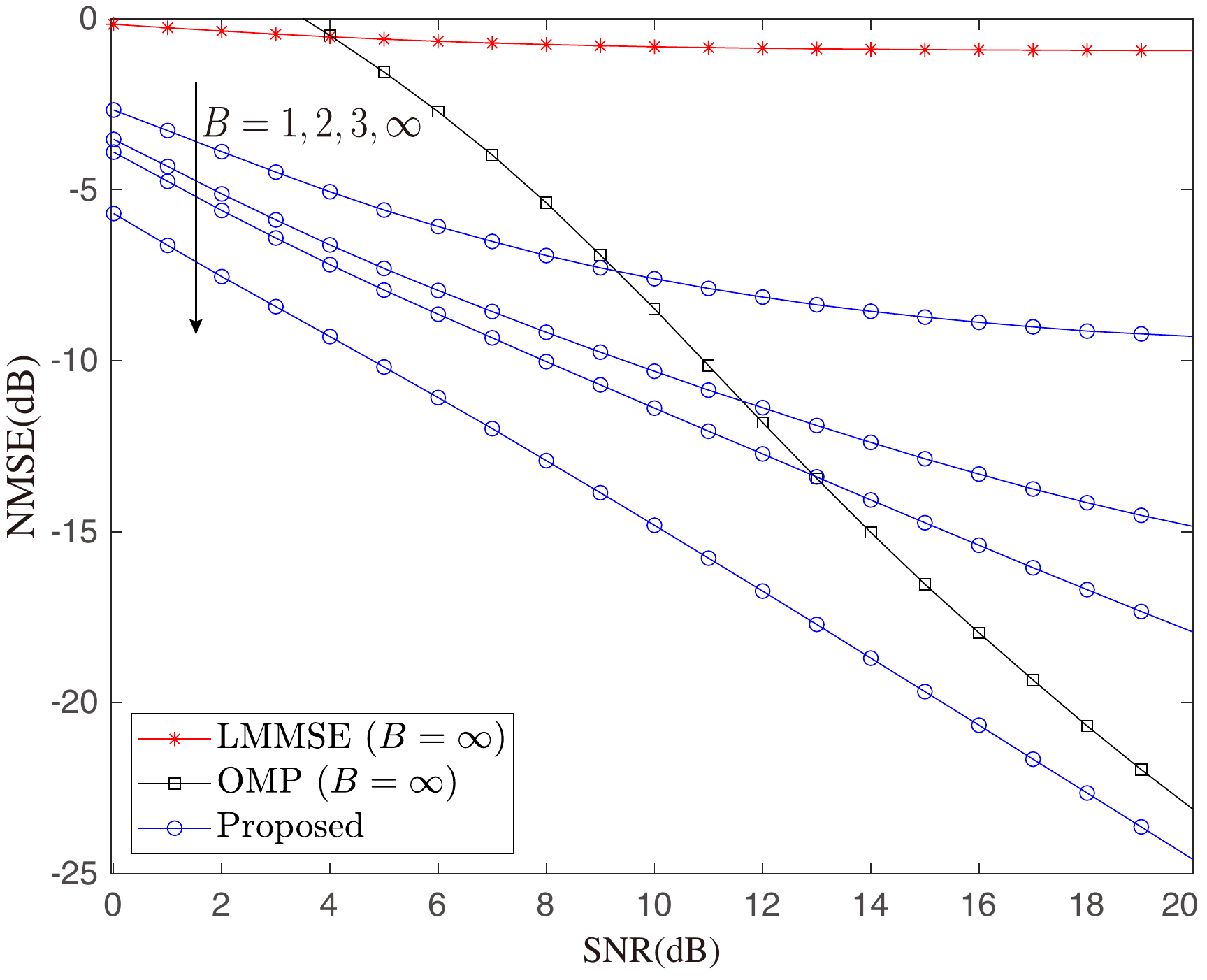}
\caption{NMSE of $\bs{X}$ versus SNR, where the dimensions of system is set as $(L,N,M,\rho)=(1024, 200, 10, 0.05)$ and $B$ refers to the quantizer's bits.
}
\label{fig:ADC}
\end{figure}

Fig.~\ref{fig:ADC} shows the figure of NMSE of $\bs{X}$ versus SNR in low-precision quantizer by varying quantizer bits. For convenience, we consider $\bs{\Sigma}_x=\sigma_x^2\bs{I}$ and $\bs{\Sigma}_w=\sigma_w^2\bs{I}$. We compare the proposed algorithm with the competing orthogonal matching pursuing (OMP) \cite{pati1993orthogonal} and classical least square (LS). As can be seen from this figure, our proposed algorithm outperforms the computing OMP especially in low SNR.

\section{Conclusion}
In this paper, we studied the generalized linear inference problem with IID row prior and row-wise mapping channel arising in many engineering regions such as wireless communications, compressed sensing, and phase retrieval. In large system limit and Bayes-optimal setting, we applied replica method from statistical mechanics to analyze the exact MMSE estimator. We presented the decouple principle, in which the input-output relation of the objective system using MMSE estimator can be decoupled into a set of single-vector additive correlated Gaussian channels. Meanwhile, the input-output mutual information relation between them was established. The explicit replica formula provides a benchmark for designing a signal estimator. To estimate $\bs{X}$ from the observation $\bs{Y}$,  we proposed a computationally efficient $\mathcal{O}(N^2+M^2N)$ message passing based algorithm using EP. By performing large system analysis, the asymptotic MSE performance of proposed algorithm can be fully tracked by its SE equations. We showed that the SE equations of proposed algorithm match perfectly the fixed point of the exact MMSE estimator predicted by replica method. That illustrates, in large system,  the proposed algorithm can attain the Bayes-optimal MSE performance if it has a unique fixed points. Finally, several numerical simulations were presented to validate the accuracy of our analytical results.

\section{Acknowledgments}
The first author is extremely grateful to Dr. Ben Grossmann, the Drexel University in Philadelphia,  for the help of matrix inverse of block matrix, Prof. Chao-Kai Wen, the National Sun Yat-sen University, for the discussion of replica symmetric assumption, and Prof. Hugo Touchette, the Stellenbosch University, for the discussion of large derivation theory.

\begin{appendices}
\section{Proof of Lemma \ref{Lemma:1}}
Let's write $\bs{Q}$ as\footnote{Some basic properties of Kronecker product can be found in \cite{hilberdink2018quasi}.}
\begin{align}
\bs{Q}
&=(\bs{I}\otimes \bs{B})\left(\bs{I}\otimes(\bs{B}^{-1}\bs{A}-\bs{I})+(\bs{1}\otimes \bs{I})(\bs{1}^{\text{T}}\otimes \bs{I})\right).
\end{align}
Then its inverse matrix is given by
\begin{align}
\nonumber
\bs{Q}^{-1}&=\left(\bs{I}\otimes(\bs{B}^{-1}\bs{A}-\bs{I})+(\bs{1}\otimes \bs{I})(\bs{1}^{\text{T}}\otimes \bs{I})\right)^{-1}\\
&\quad \times (\bs{I}\otimes \bs{B})^{-1}.
\label{Equ:A2}
\end{align}
For convenience, let's define $\bs{M}=\bs{B}^{-1}\bs{A}-\bs{I}$, then applying the Kailath Variant\footnote{
$(\bs{A}+\bs{BC})^{-1}=\bs{A}^{-1}-\bs{A}^{-1}\bs{B}(\bs{I}+\bs{CA}^{-1}\bs{B})^{-1}\bs{CA}^{-1}$.
} of Woodbury identity formula (see \cite[page 153]{bishop1995neural}), we have
\begin{align*}
\nonumber
&\left(\bs{I}\otimes \bs{M}+(\bs{1}\otimes \bs{I})(\bs{1}^{\text{T}}\otimes \bs{I})\right)^{-1}\\
\nonumber
&=(\bs{I}\otimes \bs{M})^{-1}-(\bs{I}\otimes \bs{M})^{-1}(\bs{1}\otimes \bs{I})\\
& \times \left(\bs{I}+(\bs{1}^{\text{T}}\otimes \bs{I})(\bs{I}\otimes \bs{M})^{-1}(\bs{1}\otimes \bs{I})\right)^{-1}(\bs{1}^{\text{T}}\otimes \bs{I})(\bs{I}\times \bs{M})^{-1}\\
&=\bs{I}\otimes \bs{M}^{-1}-\bs{1}\otimes \bs{M}^{-1}\left(\bs{I}+(\tau+1)\bs{M}^{-1}\right)^{-1}\bs{1}^{\text{T}}\otimes \bs{M}^{-1}
\end{align*}
Substituting the equation above into (\ref{Equ:A2}) yields
\begin{align*}
\bs{Q}^{-1}
&=\bs{I}\otimes (\bs{A}-\bs{B})^{-1}\\
&-\bs{11}^{\text{T}}\otimes ((\bs{A}-\bs{B})\bs{B}^{-1}(\bs{A}-\bs{B})+(\tau+1)(\bs{A}-\bs{B}))^{-1}.
\end{align*}

\section{}
Let's denote $\bs{E}_1=(\bs{A}-\bs{B})^{-1}$ and $\bs{E}_2=[(\bs{A}-\bs{B})\bs{B}^{-1}(\bs{A}-\bs{B})+(\bs{A}-\bs{B})]^{-1}$. From (\ref{Equ:SD1}), we have
\begin{align*}
\nonumber
\tilde{\bs{Q}}^{\star}&=-\alpha [\bs{I}\otimes \bs{E}_1-\bs{11}^{\text{T}}\otimes \bs{E}_2-\alpha (\bs{I}\otimes \bs{E}_1-\bs{11}^{\text{T}}\otimes \bs{E}_2))\\
&\qquad \times (\bs{I}\otimes (\bs{A}_z-\bs{B}_z)+\bs{11}^{\text{T}}\otimes \bs{B}_z)(\bs{I}\otimes \bs{E}_1-\bs{11}^{\text{T}}\otimes \bs{E}_2)]\\
\nonumber
&=-\alpha [\bs{I}\otimes \bs{E}_1-\bs{11}^{\text{T}}\otimes \bs{E}_2-\alpha(\bs{I}\otimes \bs{E}_1(\bs{A}_z-\bs{B}_z)\bs{E}_1+\\
&\qquad \bs{11}^{\text{T}}\otimes
(\bs{E}_1\bs{B}_z\bs{E}_1+\bs{E}_2\bs{A}_z\bs{E}_2-\bs{E}_1\bs{A}_z\bs{E}_2-\bs{E}_2\bs{A}_z\bs{E}_1)
 )].
\end{align*}
It implies
\begin{align*}
\tilde{\bs{A}}
&=-\alpha (\bs{E}_1-\bs{E}_2-\alpha (\bs{E}_1-\bs{E}_2)\bs{A}_z(\bs{E}_1-\bs{E}_2)),\\
\tilde{\bs{B}}
&=\alpha \bs{E}_2+\alpha^2(\bs{E}_1\bs{B}_z\bs{E}_1+\bs{E}_2\bs{A}_z\bs{E}_2-\bs{E}_1\bs{A}_z\bs{E}_2-\bs{E}_2\bs{A}_z\bs{E}_1).
\end{align*}

Note that $\bs{E}_2$ is symmetric.
By the fact $\bs{A}_z=\frac{1}{\alpha}\bs{A}$, we have
\begin{align*}
\tilde{\bs{A}}
&=-\alpha (\bs{E}_1-\bs{E}_2-\alpha (\bs{E}_1-\bs{E}_2)\bs{A}_z(\bs{E}_1-\bs{E}_2))\\
&=-\alpha (\bs{A}-\bs{B})^{-1}(\bs{I}-\bs{B}\bs{A}^{-1})-\alpha(\bs{A}-\bs{B})^{-1} \\
&\quad \quad \times (\bs{I}-\bs{B}\bs{A}^{-1})\bs{A}(\bs{A}-\bs{B})^{-1}(\bs{I}-\bs{B}\bs{A}^{-1})\\
&=-\alpha (\bs{A}^{-1}-\bs{A}^{-1})\\
&=\bs{0},
\end{align*}
as well as
\begin{align*}
\tilde{\bs{B}}
&=\alpha (\bs{A}-\bs{B})^{-1}\bs{BA}^{-1}+\alpha^2(\bs{A}-\bs{B})^{-1}\bs{B}_z(\bs{A}-\bs{B})^{-1}\\
& \qquad +\alpha(\bs{A}-\bs{B})^{-1}(\bs{BA}^{-1}\bs{B}-2\bs{B})(\bs{A}-\bs{B})^{-1}\\
&=\alpha^2(\bs{A}-\bs{B})^{-1}\bs{B}_z(\bs{A}-\bs{B})^{-1}-\alpha (\bs{A}-\bs{B})^{-1}\bs{B}(\bs{A}-\bs{B})^{-1}\\
&=(\bs{A}-\bs{B})^{-1}(\alpha^2 \bs{B}_z-\alpha\bs{B})(\bs{A}-\bs{B})^{-1}.
\end{align*}

\section{Derivation Using EP}
\label{App:EP}
We first simply the term $f_{n\leftarrow \ell}(\bs{x}_n)$ in (\ref{Equ:EP1}), which can be written as
\begin{align}
\nonumber
f_{n\leftarrow \ell}^{(t)}(\bs{x}_n)
&=\int \mathcal{P}(\bs{y}_{\ell}|\bs{z}_{\ell})\\
&\times \mathbb{E}\left\{\delta\left(\bs{z}_{\ell}-h_{\ell n}\bs{x}_n-\sum_{i\ne n}a_{\ell i}\bs{x}_i\right)\right\}\text{d}\bs{z}_{\ell}
\end{align}
where the expectation is taken over $\prod_{i\ne n}\mu_{i\rightarrow \ell}^{(t)}(\bs{x}_i)$. Let's define the random vector $\boldsymbol{\xi}_{i\rightarrow \ell}^{(x,t)}$ associated with $\bs{x}_i$ following $\mu_{i\rightarrow \ell}^{(t)}(\bs{x}_i)$. Based on the central limit theorem (CLT), it implies that the term $\sum_{i\ne n}a_{\ell i}\boldsymbol{\xi}_{i\rightarrow \ell}^{(x,t)}\rightarrow \mathcal{N}_c(\cdot|\sum_{i\ne n}a_{\ell i}\hat{\bs{x}}_{i\rightarrow \ell}^{(t)}, \sum_{i\ne a}|h_{\ell n}|^2\bs{Q}_{i\rightarrow \ell}^{(x,t)})$ where $\hat{\bs{x}}_{i\rightarrow \ell}^{(t)}$ and $\bs{Q}_{i\rightarrow \ell}^{(x,t)}$ are the mean and covariance matrix of $\boldsymbol{\xi}_{i\rightarrow \ell}^{(x,t)}$, respectively. Further, the term $h_{\ell n}\bs{x}_n$ is constant. Thus, we have
\begin{align}
\nonumber
&\lim_{N\rightarrow \infty}\mathbb{E}\left\{\delta\left(\bs{z}_{\ell}-h_{\ell n}\bs{x}_n-\sum_{i\ne n}a_{\ell i}\bs{x}_i\right)\right\}\\
&\rightarrow \mathcal{N}_c(\bs{z}_{\ell}|h_{\ell n}\bs{x}_n+\bs{z}_{n\leftarrow \ell}^{(t)}, \bs{Q}_{n\leftarrow \ell}^{(z,t)}),
\label{Equ:D5}
\end{align}
where
\begin{align}
\bs{z}_{n\leftarrow \ell}^{(t)}&=\sum_{i\ne n}h_{\ell i}\hat{\bs{x}}_{i\rightarrow \ell}^{(t)},\\
\bs{Q}_{n\leftarrow \ell}^{(t)}&=\sum_{i\ne n}|h_{\ell i}|^2\bs{Q}_{i\rightarrow \ell}^{(x,t)}.
\end{align}
Substituting (\ref{Equ:D5}) into (\ref{Equ:EP1}) and applying Gaussian reproduction property
\begin{align}
\nonumber
\mu_{n\rightarrow \ell}^{(t)}(\bs{x}_n)f_{n\leftarrow \ell}^{(t)}(\bs{x}_n)
&\propto \int \mathcal{P}(\bs{y}_{\ell}|\bs{z}_{\ell})\mathcal{N}_c(\bs{z}_{\ell}|\bs{z}_{\ell}^{(t)},\bs{Q}_{\ell}^{(z,t)})\\
&\times \mathcal{N}_c(\bs{x}_{\ell}|\tilde{\bs{x}}_{n\leftarrow \ell}^{(t)}(\bs{z}_{\ell}), \tilde{\bs{Q}}_{n\leftarrow \ell}^{(x,t)})\text{d}\bs{z}_{\ell},
\label{Equ:D7}
\end{align}
where the following definitions are applied
\begin{align}
\bs{z}_{\ell}^{(t)}&=\sum_nh_{\ell n}\hat{\bs{x}}_{n\rightarrow \ell}^{(t)},
\label{Equ:D6}\\
\bs{Q}^{(z,t)}_{\ell}&=\sum_n|h_{\ell n}|^2\bs{Q}_{n\leftarrow \ell}^{(x,t)},
\label{Equ:D7}\\
\tilde{\bs{Q}}_{n\leftarrow \ell}^{(x,t)}&=\left((\bs{Q}_{n\rightarrow \ell}^{(x,t)})^{-1}+|h_{\ell n}|^2(\bs{Q}_{n\leftarrow \ell}^{(z,t)})^{-1}\right)^{-1},
\label{Equ:D8}\\
\nonumber
\tilde{\bs{x}}_{n\leftarrow \ell}^{(t)}(\bs{z}_{\ell})&=\tilde{\bs{Q}}_{n\leftarrow \ell}^{(x,t)}\left((\bs{Q}_{n\rightarrow \ell}^{(x,t)})^{-1}\hat{\bs{x}}_{n\rightarrow \ell}^{(t)}\right.\\
&\left. \qquad  +h_{\ell n}^{*}(\bs{Q}_{n\leftarrow \ell}^{(z,t)})^{-1}(\bs{z}_{\ell}-\bs{z}_{n\leftarrow \ell}^{(t)})\right),
\label{Equ:D9}
\end{align}
where $(\cdot)^{*}$ denotes conjugate transpose. Using definition (\ref{Equ:D6}), we simplify (\ref{Equ:D8}) and (\ref{Equ:D9}) as
\begin{align}
\tilde{\bs{Q}}_{n\leftarrow \ell}^{(x,t)}&=\bs{Q}_{n\leftarrow \ell}^{(z,t)}(\bs{Q}_{\ell}^{(z,t)})^{-1}\bs{Q}_{n\rightarrow \ell}^{(x,t)},\\
\nonumber
\tilde{\bs{x}}_{n\leftarrow \ell}^{(t)}(\bs{z}_{\ell})&=\bs{Q}_{n\leftarrow \ell}^{(z,t)}(\bs{Q}_{\ell}^{(z,t)})^{-1}\hat{\bs{x}}_{n\rightarrow \ell}^{(t)}\\
&\qquad +h_{\ell n}^{*}\bs{Q}_{n\rightarrow \ell}^{(x,t)}(\bs{Q}_{n\rightarrow \ell}^{(z,t)})^{-1}(\bs{z}_{\ell}-\bs{z}_{n\leftarrow \ell}^{(t)}).
\end{align}
Notice that since $\bs{Q}_{n\rightarrow \ell}^{(x,t)}$ is covariance matrix which is symmetric, it implies that  $\tilde{\bs{Q}}_{n\leftarrow \ell}^{(x,t)}$ is also symmetric. The symmetric structure of $\tilde{\bs{Q}}_{n\leftarrow \ell}^{(x,t)}$ will be used in the rest of derivation.

As observed from (\ref{Equ:D7}), $\mu_{n\rightarrow \ell}^{(t)}(\bs{x}_n)f_{n\leftarrow \ell}^{(t)}(\bs{x}_n)$ is actually not a distribution. Thus, it is necessary to normalize it as a density w.r.t. $\bs{x}_n$. Taking normalization operation leads to
\begin{align}
\nonumber
&\mu_{n\rightarrow \ell}^{(t)}(\bs{x}_n)f_{n\leftarrow \ell}^{(t)}(\bs{x}_n)\\
&\propto \mathbb{E}_{\boldsymbol{\zeta}_{\ell}^{(t)}}\left\{\mathcal{N}_c(\bs{x}_{\ell}|\tilde{\bs{x}}_{n\leftarrow \ell}^{(t)}(\boldsymbol{\zeta}_{\ell}^{(t)}), \tilde{\bs{Q}}_{n\leftarrow \ell}^{(x,t)})\right\},
\end{align}
where random vector $\boldsymbol{\zeta}_{\ell}^{(t)}$ associated $\bs{z}_{\ell}$ follows $\frac{\mathcal{P}(\bs{y}_{\ell}|\bs{z}_{\ell})\mathcal{N}_c(\bs{z}_{\ell}|\bs{z}_{\ell}^{(t)}, \bs{Q}_{\ell}^{(z,t)})}{\int \mathcal{P}(\bs{y}_{\ell}|\bs{z}_{\ell})\mathcal{N}_c(\bs{z}_{\ell}|\bs{z}_{\ell}^{(t)}, \bs{Q}_{\ell}^{(z,t)})\text{d}\bs{z}_{\ell}}$. We denote its mean and covariance matrix of as
\begin{align}
\tilde{\bs{z}}_{\ell}^{(t)}&=\mathbb{E}_{\boldsymbol{\zeta}_{\ell}^{(t)}}\left\{\boldsymbol{\zeta}_{\ell}^{(t)}\right\},\\
\tilde{\bs{Q}}_{\ell}^{(z,t)}&=\text{Var}_{\boldsymbol{\zeta}_{\ell}^{(t)}}\left\{\boldsymbol{\zeta}_{\ell}^{(t)}\right\}.
\end{align}

In the sequel, we calculate the projection in (\ref{Equ:EP1})
\begin{align}
\nonumber
&\text{Proj}\left[\mu_{n\rightarrow \ell}^{(t)}(\bs{x}_n)f_{n\leftarrow \ell}^{(t)}(\bs{x}_n)\right]\\
\nonumber
&=\text{Proj}\left[\mathbb{E}_{\boldsymbol{\zeta}_{\ell}^{(t)}}\left\{\mathcal{N}_c(\bs{x}_{\ell}|\tilde{\bs{x}}_{n\leftarrow \ell}^{(t)}(\boldsymbol{\zeta}_{\ell}^{(t)}), \tilde{\bs{Q}}_{n\leftarrow \ell}^{(x,t)})\right\}\right]\\
&=\mathcal{N}_c(\bs{x}_n|\bs{x}_n^{(t)}, \bs{Q}_n^{(t)}),
\end{align}
where
\begin{align}
\nonumber
\bs{x}_n^{(t)}
&=\mathbb{E}_{\boldsymbol{\zeta}_{\ell}^{(t)}}\left\{\tilde{\bs{x}}_{n\leftarrow \ell}^{(t)}(\boldsymbol{\zeta}_{\ell}^{(t)})\right\}\\
\nonumber
&=\bs{Q}_{n\leftarrow \ell}^{(z,t)}(\bs{Q}_{\ell}^{(z,t)})^{-1}\hat{\bs{x}}_{n\rightarrow \ell}^{(t)}\\
\nonumber
&\quad +h_{\ell n}^{*}\bs{Q}_{n\rightarrow \ell}^{(x,t)}(\bs{Q}_{n\rightarrow \ell}^{(z,t)})^{-1}(\tilde{\bs{z}}_{\ell}-\bs{z}_{n\leftarrow \ell}^{(t)})\\
\nonumber
&=(\bs{Q}_{\ell}^{(z,t)}-|h_{\ell n}|^2\bs{Q}_{n\rightarrow \ell}^{(x,t)})(\bs{Q}_{\ell}^{(z,t)})^{-1}\hat{\bs{x}}_{n\rightarrow \ell}^{(t)}\\
\nonumber
&\quad +h_{\ell n}^{*}\bs{Q}_{n\rightarrow \ell}^{(x,t)}(\bs{Q}_{n\rightarrow \ell}^{(z,t)})^{-1}(\tilde{\bs{z}}_{\ell}-\bs{z}_{\ell}^{(t)}+h_{\ell n}\hat{\bs{x}}_{n\rightarrow \ell}^{(t)})\\
&=\hat{\bs{x}}_{n\rightarrow \ell}^{(t)}+h_{\ell n}^{*}\bs{Q}_{n\rightarrow \ell}^{(x,t)}(\bs{Q}_{\ell}^{(z,t)})^{-1}(\tilde{\bs{z}}_{\ell}^{(t)}-\bs{z}_{\ell}^{(t)}),\\
\nonumber
\bs{Q}_n^{(x,t)}
&=\mathbb{E}_{\boldsymbol{\zeta}^{(t)}}\left\{
\tilde{\bs{Q}}_{n\leftarrow \ell}^{(x,t)}+\tilde{\bs{x}}_{n\leftarrow \ell}^{(t)}(\boldsymbol{\zeta}_{\ell}^{(t)})\tilde{\bs{x}}_{n\leftarrow \ell}^{(t)}(\boldsymbol{\zeta}_{\ell}^{(t)})^{\dag}
\right\}\\
&\quad -\bs{x}_n^{(t)}(\bs{x}_n^{(t)})^{\dag}\label{Equ:D12}\\
\nonumber
&=\bs{Q}_{n\leftarrow \ell}^{(z,t)}(\bs{Q}_{\ell}^{(z,t)})^{-1}\bs{Q}_{n\rightarrow \ell}^{(x,t)}\\
\nonumber
&\quad +|h_{\ell n}|^2\bs{Q}_{n\rightarrow \ell}^{(x,t)}(\bs{Q}_{\ell}^{(z,t)})^{-1}\tilde{\bs{Q}}_{\ell}^{(z,t)}(\bs{Q}_{\ell}^{(z,t)})^{-1}\bs{Q}_{n\rightarrow \ell}^{(x,t)}\\
\nonumber
&=(\bs{Q}_{\ell}^{(z,t)}-|h_{\ell n}|^2\bs{Q}_{n\rightarrow \ell}^{(x,t)})(\bs{Q}_{\ell}^{(z,t)})^{-1}
\bs{Q}_{n\rightarrow \ell}^{(x,t)}\\
\nonumber
&\quad +|h_{\ell n}|^2\bs{Q}_{n\rightarrow \ell}^{(x,t)}(\bs{Q}_{\ell}^{(z,t)})^{-1}\tilde{\bs{Q}}_{\ell}^{(z,t)}(\bs{Q}_{\ell}^{(z,t)})^{-1}\bs{Q}_{n\rightarrow \ell}^{(x,t)}\\
\nonumber
&=\bs{Q}_{n\rightarrow \ell}^{(x,t)}-|h_{\ell n}|^2\bs{Q}_{n\rightarrow \ell}^{(x,t)}(\bs{Q}_{\ell}^{(z,t)})^{-1}\\
&\quad \times (\bs{Q}_{\ell}^{(z,t)}-\tilde{\bs{Q}}_{\ell}^{(z,t)})
(\bs{Q}_{\ell}^{(z,t)})^{-1}\bs{Q}_{n\rightarrow \ell}^{(x,t)}.
\label{Equ:D13}
\end{align}

Using Gaussian reproduction property, from (\ref{Equ:EP1}) we have
\begin{align}
\mu_{n\leftarrow \ell}^{(t)}(\bs{x}_n)=\mathcal{N}_c(\bs{x}_n|\hat{\bs{x}}_{n\leftarrow \ell}^{(t)}, \bs{Q}_{n\leftarrow \ell}^{(x,t)}),
\end{align}
where
\begin{align}
\!\!\!\!\!\bs{Q}_{n\leftarrow \ell}^{(x,t)}&=\left((\bs{Q}_n^{(x,t)})^{-1}-(\bs{Q}_{n\rightarrow \ell}^{(x,t)})^{-1}\right)^{-1},
\label{Equ:D10}\\
\!\!\!\!\!\hat{\bs{x}}_{n\leftarrow \ell}^{(t)}&=\bs{Q}_{n\leftarrow \ell}^{(x,t)}\left((\bs{Q}_n^{(x,t)})^{-1}\bs{x}_n^{(t)}-(\bs{Q}_{n\rightarrow \ell}^{(x,t)})^{-1}\hat{\bs{x}}_{n\rightarrow \ell}^{(t)}\right).
\label{Equ:D11}
\end{align}
Using matrix inverse lemma\footnote{
$(\bs{A}+\bs{BC})^{-1}=\bs{A}^{-1}-\bs{A}^{-1}\bs{B}(\bs{I}+\bs{CA}^{-1}\bs{B})^{-1}\bs{CA}^{-1}$.
}, (\ref{Equ:D10}) becomes
\begin{align}
\nonumber
\bs{Q}_{n\leftarrow \ell}^{(x,t)}&=\bs{Q}_n^{(x,t)}+\bs{Q}_n^{(x,t)}(\bs{Q}_{n\rightarrow \ell}^{(x,t)})^{-1}\\
&\quad \times (\bs{I}-\bs{Q}_n^{(x,t)}(\bs{Q}_{n\rightarrow \ell}^{(x,t)})^{-1})^{-1}\bs{Q}_n^{(x,t)}.
\label{Equ:D14}
\end{align}

For convenience, let's define
\begin{align}
\bs{Q}_{\text{tem}}=\bs{Q}_{n\rightarrow \ell}^{(x,t)}(\bs{Q}_{\ell}^{(z,t)})^{-1}
(\bs{Q}_{\ell}^{(z,t)}-\tilde{\bs{Q}}_{\ell}^{(z,t)})(\bs{Q}_{\ell}^{(z,t)})^{-1},
\end{align}
From (\ref{Equ:D13}), we have $\bs{Q}_n^{(x,t)}=(\bs{I}-|h_{\ell n}|^2\bs{Q}_{\text{tem}})\bs{Q}_{n\rightarrow \ell}^{(x,t)}$. Using this representation
\begin{align}
\nonumber
\bs{Q}_{n\leftarrow \ell}^{(x,t)}
&=(\bs{I}-|h_{\ell n}|^2\bs{Q}_{\text{tem}})\bs{Q}_{n\rightarrow \ell}^{(x,t)}+\frac{1}{|h_{\ell n}|^2}(\bs{I}-|h_{\ell n}|^2\bs{Q}_{\text{tem}})\\
\nonumber
&\quad \times \bs{Q}_{\text{tem}}^{-1}(\bs{I}-|h_{\ell n}|^2\bs{Q}_{\text{tem}})\bs{Q}_{n\rightarrow \ell}^{(x,t)}\\
\nonumber
&=\frac{1}{|h_{\ell n}|^2}\bs{Q}_{\text{tem}}^{-1}\bs{Q}_{n\rightarrow \ell}^{(x,t)}-\bs{Q}_{n\rightarrow \ell}^{(x, t)}\\
\nonumber
&=\frac{\bs{Q}_{\ell}^{(z,t)}
(\bs{Q}_{\ell}^{(z,t)}-\tilde{\bs{Q}}_{\ell}^{(z,t)})^{-1}\bs{Q}_{\ell}^{(z,t)}-|h_{\ell n}|^2\bs{Q}_{n\rightarrow \ell}^{(x, t)}}{|h_{\ell n}|^2}\\
&\approx \frac{1}{|h_{\ell n}|^2}\bs{Q}_{\ell}^{(z,t)}
(\bs{Q}_{\ell}^{(z,t)}-\tilde{\bs{Q}}_{\ell}^{(z,t)})^{-1}\bs{Q}_{\ell}^{(z,t)}.
\label{Equ:E1}
\end{align}
To derive $\hat{\bs{x}}_{n\leftarrow \ell}^{(t)}$, we use (\ref{Equ:D14}) rather than approximated result (\ref{Equ:E1})
\begin{align}
\nonumber
\hat{\bs{x}}_{n\leftarrow \ell}^{(t)}
&=\left[\bs{Q}_n^{(x,t)}+\bs{Q}_n^{(x,t)}(\bs{Q}_{n\rightarrow \ell}^{(x,t)})^{-1}(\bs{I}-\bs{Q}_n^{(x,t)}(\bs{Q}_{n\rightarrow \ell}^{(x,t)})^{-1})^{-1}\bs{Q}_n^{(x,t)}\right]\\
\nonumber
&\qquad \times ((\bs{Q}_n^{(x,t)})^{-1}\bs{x}_n^{(t)}-(\bs{Q}_{n\rightarrow \ell}^{(x,t)})^{-1}\hat{\bs{x}}_{n\rightarrow \ell}^{(t)})\\
\nonumber
&=(|h_{\ell n}|^2\bs{Q}_{\text{tem}})^{-1}\bs{x}_n^{(t)}-(|h_{\ell n}|^2\bs{Q}_{\text{tem}})^{-1}\hat{\bs{x}}_{n\rightarrow \ell}^{(t)}+\hat{\bs{x}}_{n\rightarrow \ell}^{(t)}\\
&=\hat{\bs{x}}_{n\rightarrow \ell}^{(t)}+\frac{h_{\ell n}^{*}}{|h_{\ell n}|^2}\bs{Q}_{\ell}^{(z,t)}(\bs{Q}_{\ell}^{(z,t)}-\tilde{\bs{Q}}_{\ell}^{(z,t)})^{-1}(\tilde{\bs{z}}_{\ell}^{(t)}-\bs{z}_{\ell}^{(t)}).
\label{Equ:E2}
\end{align}

For convenience, we define
\begin{align}
\bs{Q}_{\ell}^{(s,t)}&=(\bs{Q}_{\ell}^{(z,t)})^{-1}(\bs{Q}_{\ell}^{(z,t)}-\tilde{\bs{Q}}_{\ell}^{(z,t)})(\bs{Q}_{\ell}^{(z,t)})^{-1},\\
\bs{s}_{\ell}^{(t)}&=(\bs{Q}_{\ell}^{(z,t)})^{-1}(\tilde{\bs{z}}_{\ell}-\bs{z}_{\ell}^{(t)}).
\end{align}
Using the definitions above to represent $\bs{Q}_{n\leftarrow \ell}^{(x,t)}$ in (\ref{Equ:E1}) and $\hat{\bs{x}}_{n\leftarrow \ell}^{(t)}$ in (\ref{Equ:E2}), we obtain
\begin{align}
\!\!\!\!\!\!
\bs{Q}_{n\leftarrow \ell}^{(x,t)}&=(|h_{\ell n}|^2\bs{Q}_{\ell}^{(s,t)})^{-1},\\
\!\!\!\!\!\!\hat{\bs{x}}_{n\leftarrow \ell}^{(t)}&=(|h_{\ell n}|^2\bs{Q}_{\ell}^{(s,t)})^{-1}(|a_{\ell}|^2\bs{Q}_{\ell}^{(s,t)}\hat{\bs{x}}_{n\rightarrow \ell}^{(t)}+h_{\ell n}^{*}\bs{s}_{\ell}^{(t)}).
\end{align}
Up to this point, we have solved the equation (\ref{Equ:EP1}). Then, we move to the calculation of (\ref{Equ:EP2}). Also by Gaussian reproduction property, the product of $L$ Gaussian terms with the same argument is proportional to a Gaussian vector:
\begin{align}
\prod_{\ell }\mu_{n\leftarrow \ell}^{(t)}(\bs{x}_n)\propto \mathcal{N}_c(\bs{x}_n|\bs{r}_n^{(t)}, \bs{Q}_n^{(r,t)}),
\end{align}
where
\begin{align}
\bs{Q}_n^{(r,t)}&=\left(\sum_{\ell=1}^L |h_{\ell n}|^2\bs{Q}_{\ell}^{(s,t)}\right)^{-1},\\
\bs{r}_n^{(t)}&=\bs{Q}_n^{(r,t)}\sum_{\ell=1}^L (|h_{\ell n}|^2\bs{Q}_{\ell}^{(s,t)}\hat{\bs{x}}_{n\rightarrow \ell}^{(t)}+h_{\ell n}^{*}\bs{s}_{\ell}^{(t)}).
\label{Equ:E4}
\end{align}
Let's define a random vector $\boldsymbol{\xi}_n^{(t)}$ associated with $\bs{x}_n$ following $\frac{\mathcal{P}_{\textsf{X}}(\bs{x}_n)\mathcal{N}_c(\bs{x}_n|\bs{r}_n^{(t)}, \bs{Q}_n^{(r,t)})}{\int \mathcal{P}_{\textsf{X}}(\bs{x}_n)\mathcal{N}_c(\bs{x}_n|\bs{r}_n^{(t)}, \bs{Q}_n^{(r,t)})\text{d}\bs{x}_n}$. Then the projection in (\ref{Equ:EP2}) is given by
\begin{align}
\text{Proj}\left[\mathcal{P}(\bs{x}_n)\prod_{\ell=1}^L\mu_{n\leftarrow l}^{(t)}(\bs{x}_n)\right]
=\mathcal{N}_c(\bs{x}_n|\hat{\bs{x}}_n, \hat{\bs{Q}}_n^{(x,t)}),
\end{align}
where $\hat{\bs{x}}_n^{(t)}$ and $\hat{\bs{Q}}_n^{(x,t)}$ denote the mean and covariance matrix of $\boldsymbol{\xi}_n^{(t)}$.

Also by Gaussian reproduction property, from (\ref{Equ:EP2})
\begin{align}
\hat{\bs{Q}}_{n\rightarrow \ell}^{(x,t+1)}&\approx \hat{\bs{Q}}_{n}^{(x,t+1)},\\
\hat{\bs{x}}_{n\rightarrow \ell}^{(t+1)}&\approx \hat{\bs{x}}_n^{(t+1)}-h_{\ell n}^{*}\hat{\bs{Q}}_n^{(x,t+1)}\bs{s}_{\ell}^{(t)}.
\label{Equ:E3}
\end{align}
Taking the equations above into $\bs{Q}_{\ell}^{(z,t)}$ in (\ref{Equ:D6}) and $\bs{z}_{\ell}^{(t)}$ in (\ref{Equ:D7}) yields
\begin{align}
\bs{Q}_{\ell}^{(z,t)}&=\sum_{n=1}^N |h_{\ell n}|^2\hat{\bs{Q}}_n^{(x,t)},\\
\bs{z}_{\ell}^{(t)}&=\sum_{n=1}^N h_{\ell n}\hat{\bs{x}}_n^{(t)}-\bs{Q}_{\ell}^{(z,t)}\bs{s}_{\ell}^{(t-1)}.
\end{align}
Substituting (\ref{Equ:E3}) into (\ref{Equ:E4}) yields
\begin{align}
\bs{r}_n^{(t)}
&\approx \hat{\bs{x}}_n^{(t)}+\bs{Q}_n^{(r,t)}\sum_{\ell=1}^Lh_{\ell n}^{*}\bs{s}_{\ell}^{(t)}.
\end{align}
Totally, combing those procedures finishes the derivation of proposed algorithm using EP projection.

\end{appendices}

\bibliographystyle{IEEEtran}%

\bibliography{ZQY_bib}

\end{document}